\documentclass[a4paper]{article}
\usepackage[margin=1.2in]{geometry}
\usepackage{amsmath}
\usepackage{algorithmic}
\usepackage{algorithm}
\usepackage{amsthm}
\usepackage{amsfonts}
\usepackage{comment}
\usepackage{graphicx}
\usepackage{hyperref}
\usepackage{color}
\usepackage{algorithm,algorithmic}

\usepackage{tabularx}
\usepackage{bigstrut}
\usepackage{amsmath, amssymb, multirow, paralist}
\usepackage{algorithm,algorithmic}
\usepackage{graphicx}
\usepackage{float}
\usepackage{subfigure}
\usepackage{color}
\usepackage[latin1]{inputenc}
\usepackage{hyperref}

\usepackage{fourier}
 \usepackage[T1]{fontenc}

\definecolor{mycolor}{RGB}{135,0,130}

\hypersetup{
colorlinks=true,
citecolor=mycolor,
linkbordercolor={1 1 1}, 
citebordercolor={1 1 1} 
}

\newcommand{\ct}[1] {\textcolor{black}{#1}}

\def \D {\mathcal{D}}
\def \N {\mathcal{N}}
\def \L {\mathcal{L}}
\def \R {\mathbb{R}}

\def \U {\mathcal{U}}
\def \I {\mathcal{I}}
\def \Ub {\mathbf{U}}
\def \Vb {\mathbf{V}}
\def \Rb {\mathbf{R}}
\def \tr {{\rm{Tr}}}
\def \Cb {\mathbf{C}}
\def \Sb {\mathbf{S}}
\def \Wb {\mathbf{W}}

\newtheorem{remark}{Remark}

\usepackage[mathscr]{euscript}

\title{Matrix Factorization with  Explicit Trust and Distrust Relationships}

\author{Rana Forsati \\
\small{Shahid Beheshti University, G.C., Tehran, Iran} \\ 
\small{\texttt{r\_forsati@sbu.ac.ir}}
\and Mehrdad Mahdavi\\  
\small{Michigan State University, Michigan, USA} \\ 
\small{\texttt{mahdavim@cse.msu.edu}}
\and Mehrnoush Shamsfard\\  
\small{Shahid Beheshti University, G.C., Tehran, Iran} \\ 
\small{\texttt{m\_shams@sbu.ac.ir}}
\and Mohamed Sarwat\\  
\small{University of Minnesota, Minneapolis, USA} \\ 
\small{\texttt{sarwat@cs.umn.edu}}
}
\date{}
\begin{document}

\maketitle

\begin{abstract}
With the advent of online social networks,  recommender  systems have became crucial  for the success of many online applications/services due to their  significance role in  tailoring these applications  to user-specific needs or preferences.  Despite their increasing popularity, in general  recommender systems suffer from  the \textit{data sparsity}  and the \textit{cold-start}  problems. To alleviate these issues, in recent years there has been an upsurge of interest in exploiting social information such as trust relations among users along with the rating data to improve the performance of recommender systems. The main motivation for exploiting trust information in recommendation process stems from the observation that the ideas we are exposed to and the choices we make are significantly influenced by our social context. However, in large user communities, in addition to trust relations, the distrust  relations  also exist between users. For instance, in Epinions the concepts of  personal "web of trust"  and personal "block list" allow  users to categorize their friends based on the quality of reviews into trusted and distrusted friends, respectively.  Hence,  it will be interesting to incorporate this new source of information  in recommendation as well. In contrast to the incorporation of trust  information  in recommendation which is thriving, the potential of explicitly  incorporating distrust relations is almost unexplored.  In this paper, we  propose a matrix factorization based model for recommendation in social rating networks that properly incorporates both  trust and distrust relationships  aiming to improve the quality of  recommendations and mitigate the data sparsity and the cold-start users issues.  Through experiments on the Epinions data set, we show that our new algorithm outperforms its standard trust-enhanced or distrust-enhanced counterparts with respect to accuracy, thereby demonstrating the positive effect that  incorporation of explicit distrust information can have on  recommender systems.

\end{abstract}

\section{Introduction}\label{sec:intro}
The huge amount  of information available on the Web has made it increasingly challenging to cope with this information overload and  find the most relevant information one is really interested in. Recommender systems  intend to provide users with recommendations of products they
might appreciate, taking into account their past ratings, purchase history, or interest.  The recent proliferation of online social networks have further enhanced the need for such systems.  Therefore, it is obvious why  such  systems are indispensable for the success of many online applications such as Amazon, iTunes and Netflix to guide the search process and help users to effectively find the information or products  they are looking for~\cite{miller2004pocketlens}. Roughly speaking, the overarching goal of recommender systems is to identify a subset of items (e.g. products, movies, books, music, news, and web pages) that are likely to be more interesting to users based on their interests~\cite{deshpande2004item,204,forsati2010effective,bobadilla2013recommender}.

In general, most widely used recommender systems (RS)  can be broadly classified into  content-based (CB), collaborative filtering (CF), or hybrid methods~\cite{100}. In CB  recommendation, one tries to recommend items similar to those a given user preferred in the past. These methods usually rely on the   external information such as explicit item descriptions, user profiles, and/or the appropriate features extracted from items to analyze item similarity or user preference to provide recommendation.  In contrast, CF recommendation, the most popular method adopted by contemporary recommender systems, is based on the core assumption that similar users on similar items express similar interest, and it usually relies on the rating information to build a \textit{model} out of the rating information in the past without having access to  external information required in CB methods. The  hybrid approaches were  proposed  that combine both CB and CF based recommenders to gain advantages  and avoid certain limitations of each type of  systems~\cite{good1999combining,soboroff1999combining,pazzani1999framework,melville2002content,pavlov2002maximum,talabeigi2010hybrid,forsati2013fuzzy}.

The essence of CF lies in analyzing the neighborhood information of past users and items' interactions in the user-item rating matrix to generate personalized recommendations based on the preferences of other users with similar behavior.  CF has been shown to be an effective approach to recommender systems. The advantage of these types of recommender systems over content-based  RS is that the CF based methods do not require  an explicit  representation of the items in terms of features,  but it is based only on the judgments/ratings of the users. These CF algorithms are mainly divided into two main categories~\cite{205}:  \textit{memory-based} methods (also known as neighborhood-based methods)~\cite{1100,1113} and \textit{model-based} methods~\cite{1117,1118,srebro2003weighted,1112}. Recently,  another direction in CF considers how to   combine memory-based and model-based approaches to take advantage of both types of methods, thereby building a more accurate hybrid recommender system~\cite{pennock2000collaborative,xue2005scalable,koren2008factorization}.

The heart of memory-based CF methods is the measurement of similarity based on ratings of items given by users: either the similarity of users (user-oriented CF)~\cite{1005}, the similarity of items (items-oriented CF)~\cite{sarwar2001item}, or combined user-oriented
and item-oriented collaborative filtering approaches to overcome the
limitations specific to either of them~\cite{wang2006unifying}. The user-oriented CF computes the similarity among users, usually based on user profiles or past behavior, and seeks consistency in the predictions among similar users~\cite{1007,1117}. The item-oriented CF, on the other hand, allows input of additional item-wise information and is also capable of capturing the interactions among them.  If the rating of an item by a user is {unavailable}, collaborative-filtering methods estimate it by computing a weighted average of known ratings of the items from the most similar users.

Memory-based collaborative filtering is most effective when users have expressed enough ratings to have common ratings with other users, but it performs poorly for so-called {\it cold-start} users. Cold-start users are new users who have expressed only a few ratings. Thus, for memory based CF methods to be effective, large amount of user-rating data are required.  Unfortunately, due to the sparsity of the user-item rating matrix, memory-based methods may fail to correctly identify the most similar users or items, which in turn decreases  the recommender accuracy. Another major issue that memory-based methods suffer from is the scalability problem.  The reason is essentially the fact that  when the number of users and items are very large, which is common in many real world applications,   the search to identify $k$ most similar neighbors of the active user is computationally burdensome.  In summary, data sparsity and non-scalability issues are two main issues current memory based methods suffer from. 

To overcome the limitations of memory-based methods, model-based approaches have been proposed, which establish a model using the observed ratings that can interpret the given data and predict the unknown ratings~\cite{100}.   In contrast to the memory-based algorithms, model-based algorithms try to model the users based on their past ratings
and use these models to predict the ratings on
unseen items. In model-based CF the goal is to employ statistical and machine learning techniques to learn models from the data and make recommendations based on the learned model. Methods in this category include aspect model~\cite{1117,1118}, clustering methods~\cite{1999}, Bayesian model~\cite{2007}, and low dimensional linear factor models such as matrix factorization (MF)~\cite{010,srebro2003weighted,1112,1114}.  Due to its efficiency in handling very huge data sets, matrix factorization based methods have become one of the most popular models among the model-based methods, e.g. weighted low rank matrix factorization~ \cite{srebro2003weighted}, weighted nonnegative matrix factorization (WNMF) ~\cite{1112}, maximum margin matrix factorization (MMMF)~\cite{010} and probabilistic matrix factorization (PMF)~\cite{1114}. These methods assume that user preferences can be modeled by only a small number of latent factors~\cite{2012} and  all focus on fitting the user-item rating matrix using low-rank approximations only based on the observed  ratings. The recommender system we will propose in this paper adhere to the model-based factorization paradigm. 

Although latent factor models and in particular matrix factorization are able to generate high quality recommendations, these techniques also suffer from the data sparsity problem in real-world scenarios and fail to address  users who rated only a few items. For instance, according to~\cite{sarwar2001item}, the density of non-missing ratings in most commercial recommender systems is less than one or even much less. Therefore, it is unsatisfactory  to rely predictions  on such small amount of data which  becomes more challenging in the presence of large number of users or items. This  observation necessitates  tackling the data sparsity problem in an affirmative manner  to be able to generate more accurate recommendations.

One of the most prominent  approaches to tackle the data sparsity problem is to compensate for the lack of information in rating matrix with other sources of side information which are available to the recommender system.  For example, social media applications allow users to connect with each other and to interact with items of interest such as songs, videos,  pages, news, and groups.  In such networks the ideas we are exposed to and the choices we make are significantly influenced by our social context. More specifically, users generally tend to connect with other users due to some commonalities they share, often reflected in similar interests. Moreover, in many real-life applications it may be the case that only social information about certain users is available while interaction data between the items and those users has not yet been observed. Therefore, the social data accumulated in social networks  would be a rich source of information for the recommender system to utilize as  side information to alleviate the data sparsity problem. To accomplish this goal, in recent years the trust-based recommender systems  became an emerging field to provide users personalized item recommendations based on the historical ratings given by users  and the trust relationships among users (e.g., social friends).

Social-enhanced recommendation  systems are becoming of greater significance and practicality with the increased availability of online reviews, ratings, friendship links, and follower relationships. Moreover,  many e-commerce  and consumer review websites provide both reviews of products and a social network structure among the reviewers. As an example, the e-commerce site \texttt{Epinions}~\cite{Guha-2004-propagation} asks its users to indicate which reviews/users they trust and use these trust information to rank the reviews of products. Similar patterns can be found in online communities such as \texttt{Slashdot} in which millions of users post news and comment daily and are capable of tagging other users as friends/foes or fans/freaks.  Another  example is the ski  mountaineering site \texttt{Moleskiing}~\cite{avesani2005trust} which enables users to share their opinions about the snow conditions of the different ski routes and also express
how much they trust the other users.  Another well-known example is the \texttt{FilmTrsut} system~\cite{golbeck2006filmtrust}, an online social network that provides  movie rating and review features to its users. The social networking component of the website requires users to provide a trust rating for each person they add as a friend. Also users on \texttt{Wikipedia} can vote for or against the nomination of others to adminship~\cite{burke2008mopping}. These websites have come to play an important role in guiding users' opinions on products and in many cases also influence their decisions in buying or not buying the product or service. The results of experiments in~\cite{298} and of similar works confirm  that a social network can be exploited to improve the quality of recommendations. From this point of view, traditional recommender systems that ignore the social  structure between users may no longer be suitable.

A fundamental assumption in social based recommender systems which has been adopted by almost all of the relevant literature is  that if two users have friendship relation, then the recommendation from his or her friends probably has higher trustworthiness than strangers. Therefore the goal becomes how to combine  the user-item rating matrix with the social/trust network of a user to boost the accuracy of recommendation system and alleviate the sparsity problem.  Over the years, several studies have addressed the issue of the transfer of trust among users in online social networks. These studies exploit the fact that trust can be passed from one member to another in a social network, creating trust chains, based on its propagative and transitive  nature~\footnote{We note  that while the concept of trust has been studied in many disciplines including sociology, psychology, economics, and computer science from different perspectives,  but the issue of  propagation and transitivity have often been debated  in literature and  different authors have reached  different conclusions (see for example~\cite{sherchan2013survey} for a thorough discussion)}. Therefore,  some recommendation methods fusing social relations by regularization~\cite{60,61,62,zhu2011social} or factorization~\cite{64,150,1114,0002,srebro2003weighted,0003,0001} were proposed that exploit the trust relations in the social network.

Also, the results of incorporating the trust information in recommender systems is appealing and has been the focus of many researchers in the last few years,  but, in large user communities, besides the trust relationship between users, the distrust relationships are also unavoidable.  For example, Epinions  provided the feature  that enables  users to categorize other users in a personal \textit{web of trust} list based on their quality as a reviewer. Later on, this feature integrated with the  concept of personal \textit{block list}, which reflects the members that are  distrusted by a particular user. In other words, if a user encounters a member whose reviews are consistently offensive, inaccurate, or otherwise low quality, she can add that member to her block list. Therefore, it would be tempting to investigate  whether or not distrust information could be effectively  utilized  to boost the accuracy  of recommender systems as well.

In contrast to trust information for which there has been a great research,  the potential advantage/disadvantage of explicitly utilizing  distrust information is almost unexplored. Recently, few attempts have been made to explicitly incorporate the distrust relations  in recommendation process~\cite{Guha-2004-propagation,44444,VictorCCT11,33333}, which demonstrated that the recommender systems can benefit from the proper incorporation of distrust relations in social networks.  However, despite these positive results, there are some unique challenges involved in distrust-enhanced recommender systems.  In particular, it has  proven challenging to model distrust propagation in a manner which is both logically consistent and psychologically plausible. Furthermore, the naive modeling of distrust as negative trust raises a number of challenges- both algorithmic and philosophical.  Finally, it is an open challenge  how to incorporate trust and distrust relations in model-based methods simultaneously. This paper  is concerned with these  questions and gives an affirmative solution to challenges involved with distrust-enhanced recommendation. In particular, the proposed method  makes it possible to simultaneously incorporate both trust and distrust relationships in  recommender systems to increase the prediction accuracy.  To the best of our knowledge, this is the first work that  models distrust relations into the matrix factorization problem along with trust relations at the same time. 

The main intuition behind the proposed algorithm is that one can interpret the distrust relations between users as the \textit{dissimilarity} in their preferences. In particular, when a user $u$ distrusts another user $v$, it  indicates that user $u$ disagrees with most of the opinions issued, or ratings made by user $v$. Therefore, the latent features of user $u$ obtained by matrix factorization must be as dissimilar as possible to $v$'s latent features.  In other words, this intuition suggests to directly incorporate the distrust into recommendation by considering distrust  as reversing the deviation of latent features. However, when combined with the trust relations between users, due to  the contradictory role  of trust and distrust relations  in propagating social information in the  matrix factorization process, this idea fails to effectively capture  both relations simultaneously. This statement also follows from  the preliminary experimental results in~\cite{VictorCCT11} for memory-based CF methods that demonstrated regarding distrust as an indication to reverse deviations in not the right way to incorporate distrust.

To remedy this problem, we settle to a less ambitious goal and propose another  method to facilitate the learning from both types of relations. In particular, we try to learn latent features in a manner that the latent features of users who are distrusted by the user $u$  have a guaranteed minimum  dissimilarity  gap from the worst dissimilarity of users who are trusted by user $u$.  By this formulation, we ensure that when user $u$ agrees on an item with one of his trusted friends,  he/she will disagree on the same item with his distrusted friends with a minimum predefined margin.  We note that this idea significantly departs from the existing works in distrust-enhanced memory based recommender systems~\cite{VictorCCT11,33333}, that employ the distrust relations to either \textit{filter} out or \textit{debug} the trust relations to reduce the prediction task to a trust-enhanced recommendation.  In particular, the proposed method ranks the latent features of trusted and distrusted friends of each user to reflect the effect  of relation in factorization.

\paragraph{Summary of Contributions}{This work  makes the following key contributions:

\begin{itemize}
\item A matrix factorization based algorithm for simultaneous incorporation of trust and distrust relationships in recommender systems. To the best of our knowledge, this is the first model-based recommender algorithm that is able to leverage both types of relationships in recommendation.
\item An efficient stochastic optimization algorithm to solve the  optimization problem which makes the proposed method scalable to large social networks. 
\item An empirical investigation of the consistency of the social relationships with rating information. In particular, we examine to what extent trust and distrust relations between users are aligned with the ratings they issued on items. 
\item An exhaustive set of experiments on Epinions data set to empirically evaluate the performance of the proposed algorithm and demonstrate its merits and advantages.

\item A detailed comparison of the proposed algorithm to the state-of-the-art trust/distrust enhanced memory/model based recommender systems. 
\end{itemize}
}
 \paragraph{Outline}{The rest of this paper is organized as follows. In Section~\ref{sec:related}  we draw connections to and put our work in context of some of the most recent work on social recommender systems. Section~\ref{sec:mf} formally  introduces the matrix factorization problem, an optimization based framework to solve it, and its extension to incorporate the trust relations between users. The proposed algorithm along with optimization methods are discussed in Section~\ref{sec:mftrust-distrust}.  Section~\ref{sec:exp} includes our experimental result on Epinions data set which demonstrates the merits of the proposed algorithm in alleviating data sparsity problem in rating matrix and generating more accurate recommendations. Finally, Section~\ref{sec:conc-future} concludes the paper and discusses few directions as future work. }

\section{Related Work on Social Recommendation}\label{sec:related}
Earlier in the introduction, we discussed some of the main lines of research on recommender system; here, we survey further lines of study that are  most directly related to our work on social-enhanced recommendation.   Many successful algorithms have been developed over the past few years to incorporate social information in recommender systems.  After reviewing trust-enhanced memory-based approaches, we discuss some model-based approaches for recommendation in social networks with trust  relations. Finally,  we review  major approaches in distrust modeling and distrust-enhanced recommender systems.
 
\subsection{Trust Enhanced Memory-based  Recommendation}{Social network data has been widely investigated in the memory-based approaches. These methods typically explore the social network and find a neighborhood of users trusted (directly or indirectly) by a user and perform the recommendation by aggregating their ratings. These methods use the transitivity of trust and propagate trust to indirect neighbors in the social network~\cite{20004,massa2009trust,2009,900,60,403,koren2009matrix}.

In~\cite{20004}, a trust-aware collaborative filtering method for recommender systems is proposed. In this work, the collaborative filtering process is informed by the reputation of users, which is computed by propagating trust.~\cite{2009} proposed a method   based on the random walk  algorithm to utilize social connection and other social annotations to improve recommendation accuracy. However, this method does not utilize the rating information and is not applicable to constructing a random walk graph in real data sets. TidalTrust~\cite{golbeck2006generating} performs a modiÞed breadth first search in the trust network to compute a prediction. To compute the trust value between user $u$ and $v$ who are not directly connected, TidalTrust aggregates the trust value between $u$'s direct neighbors and $v$ weighted by the direct trust values of $u$ and its direct neighbors. 

{MoleTrust}~\cite{20004,massa2005controversial,2007} does the same idea as TidalTrust, but  MoleTrust considers all the  raters up to a fixed maximum-depth given as an input, independent of any specific user and item. The trust metric in MoleTrust consists of two major steps. First, cycles in trust networks are removed. Therefore, removing trust cycles beforehand from trust networks can significantly speed up the proposed algorithm because every user only needs to be visited once to infer trust values. Second, trust values are calculated based on the obtained directed acyclic graph by performing a simple graph random walk:

TrustWalker~\cite{900} combines trust-based and item-based recommendation to consider enough ratings without suffering from noisy data. Their experiments show that TrustWalker outperforms other existing memory based approaches.  Each random walk on the user trust graph returns a predicted rating for user $u$ on target item $i$. The probability of stopping is directly proportional to the similarity between the target item and the most similar item $j$, weighted by the sigmoid function of step size $k$. The more the similarity, the greater the probability of stopping and using the rating on item $j$ as the predicted rating for item $i$. As the step size increases, the probability of stopping decreases. Thus ratings by closer friends on similar items are considered more reliable than ratings on the target item by friends further away. 

 We note that all these methods are  neighborhood-based methods which employ only heuristic algorithms to generate recommendations. There are several problems with this approach. The relationship between the trust network and the user-item matrix has not been studied systematically. Moreover,  these methods are not scalable to very large data sets since they may need to calculate the pairwise user similarities and pairwise user trust scores.
}

\subsection{Trust Enhanced Model-based  Recommendation} Recently, researchers exploited matrix factorization techniques to learn latent features for users and items from the observed ratings and fusing social relations among users with rating data as will be detailed in Section~\ref{sec:mf}. These methods can be divided into two types: regularization-based methods and factorization-based methods. Here we review some existing matrix factorization algorithms that incorporate trust information in the factorization process.

\subsubsection{Regularization based Social Recommendation}

Regularization based methods typically add regularization term to the loss function and minimize it. Most recently, Ma~\cite{62} proposed an idea based on social regularized matrix factorization to make recommendation based on social network information. In this approach, the social regularization term is added to the loss function, which measures the difference between the latent feature vector of a user and those of his friends. 
The probability model similar to the model in~\cite{62} is proposed by Jamali~\cite{60}. The graph Laplacian regularization term of social relations is added into the loss function in~\cite{61} and minimizes the loss function by alternative projection algorithm. Zhu et a l.~\cite{zhu2011social} used the same model in~\cite{61} and built graph Laplacian of social relations using three kinds of kernel functions. In~\cite{0099}, the minimization problem is formulated as a low-rank semidefinite optimization problem.

\subsubsection{Factorization based Social Recommendation} 
In factorization-based methods, social relationship between users  are represented as social relation matrix, which is factored as well as the rating matrix. The loss function is the weighted sum of the social relation matrix factorization error and the rating matrix factorization error. For instance, SoRec~\cite{64} incorporates the social network graph into probabilistic matrix factorization model by simultaneously factorizing the user-item rating matrix and the social trust networks by sharing a common latent low-dimensional user feature matrix~\cite{0099}. The experimental analysis shows that this method generates better recommendations than the non-social  filtering algorithms~\cite{403}. However, the disadvantage of this work is that although the usersÕ social network is integrated into the recommender systems by factorizing the social trust graph, the real world recommendation processes are not reflected in the model. Two sets of different feature vectors are assumed for users which makes the interpretability of the model very hard~\cite{403,55555}. This drawback not only causes lack of interpretability in the model, but also affects the recommendation qualities. A better model named Social Trust Ensemble (STE)~\cite{55555} is proposed by the same authors, by making the latent features of a user's direct neighbors affect the rating of the user. Their method is a linear combination of basic matrix factorization approach and a social network based approach.  Experiments show that their model outperforms the basic matrix factorization based approach and existing trust based approaches. However, in their model, the feature vectors of direct neighbors of $u$ affect the ratings of $u$ instead of affecting the feature vector of $u$.  This model does not handle trust propagation. 
Another method for recommendation in social networks has been proposed in~\cite{44444}. This method is not a generative model and  defines a loss function  to be minimized. The main disadvantage of this method is that it punishes the users with lots of social relations more than other users. 
Finally, SocialMF~\cite{403} is a matrix factorization based model which incorporates social influence by making the features of every user depend on the features of his/her direct neighbors in the social network. 


\subsection{Distrust Enhanced Social Recommendation}
In  contrast to incorporation of trust relations, unfortunately most of the literature on social recommendation totally ignore the potential of distrust information in boosting the accuracy of recommendations. In particular, only recently few work started to investigate  the rule of distrust information in recommendation process both from theoretical and empirical viewpoints~\cite{Guha-2004-propagation,ziegler2005propagation,nurri-2008,ziegler2009propagating,44444,wierzowiecki2010efficient,VictorCCT11,victor2011practical,verbiest2012trust,33333}. Although these studies  have  shown that distrust information can be plentiful, but there is a significant gap in clear understanding of  distrust in recommender systems. The most important reasons for this shortage are the lack of data sets that contain distrust information and dearth of a unified consensus on modeling and propagation of distrust.

A formal framework of trust propagation schemes, introducing the formal and computational treatment of distrust propagation has been developed in~\cite{Guha-2004-propagation}. In an extension of this work,~\cite{ziegler2009propagating} proposed clever adaptations in order to handle distrust and sinks such as trust decay and normalization. In~\cite{wierzowiecki2010efficient},  a trust/distrust propagation algorithm called CloseLook is proposed, which is capable of using the same kinds of trust propagation as the algorithm proposed by~\cite{Guha-2004-propagation}.~\cite{leskovec2010predicting} extended the results  by~\cite{Guha-2004-propagation} using a machine-learning framework (instead of the propagation algorithms based on an adjacency matrix) to enable the evaluation of the most informative structural features for the prediction task of positive/negative links in online social networks.  A  comprehensive framework that computes trust/distrust estimations for user pairs in the network using trust metrics is build in~\cite{victor2011practical}: given two users in the trust network, we can search for a path between them and propagate the trust scores along this path to obtain an estimation. When more than one path is available, we may single out the most relevant ones (selection), and aggregation operators can then be used to combine the propagated trust scores into one final trust score, according  to  different  trust score propagation operators.

\cite{44444} was the first seminal work to demonstrate that the incorporation of distrust information could be beneficial based on a model-based recommender system.  In~\cite{victor2011practical} and~\cite{33333} the same question is addressed in memory-based approaches. In particular,~\cite{33333} embarked upon the distrust-enhanced recommendation and showed that with careful incorporation of distrust metric, distrust-enhanced recommender systems are able to outperform their trust-only counterparts. The main rational behind the algorithm proposed in~\cite{33333} is to employ the distrust information to debug or filter out the users' propagated web of trust. It is also has been realized that the debugging methods must exhibit a moderate behavior in order to be effective. ~\cite{verbiest2012trust} addressed the problem of considering the length of the paths that connect two users for computing trust-distrust between them, according to the concept of \textit{trust decay}. This work also introduced several aggregation strategies for trust scores with variable path lengths

Finally we note that the aforementioned works try to either model or utilize the trust/distrust information. In recent years there has been an upsurge of interest in predicting the trust and distrust relations in a social network~\cite{leskovec2010predicting,dubois2011predicting,bachi2012classifying,patil2013quantifying}. For instance,~\cite{leskovec2010predicting} casts the problem as a sign prediction problem (i.e., +1 for friendship and -1 for opposition) and utilizes machine learning methods to predict the sign of links in the social network. In~\cite{dubois2011predicting}  a new method is presented for computing both trust and distrust by combining an inference algorithm that relies on a probabilistic interpretation of trust based on random graphs with a modified spring-embedding algorithm to classify an edge.  Another direction of research is to  examine the consistency of social relations with theories in social psychology~\cite{cartwright1956structural,leskovec2010signed}. Our work significantly departs from  these works on prediction or consistency analysis of social relations, and aims  to effectively incorporate the distrust information in matrix factorization for effective recommendation.

\begin{table}[t]
\begin{center}
\caption{Summary of notations consistently used in the paper and their meaning.\label{table:notations}}{%
\begin{tabular} {| l || l |  }
\hline 
Symbol & Meaning\\
\hline \hline 
$\U = \{u_1, \cdots, u_n\}$, $n$ & The set of users in system and the number of users\\
\hline 
$\I = \{i_1, \cdots, i_m\}$, $m$ & The set of items and the number of items\\
\hline 
$k$& The dimension of latent features   in factorization \\
\hline
$\Rb \in \R^{n \times m}$ & The partially observed rating matrix \\
\hline 
$\Omega_\Rb$, $|\Omega_\Rb|$ & The set of observed entires in rating matrix and its size \\
\hline 
$\Ub \in \R^{n \times k}$&The matrix of  latent features for users\\
\hline 
$\Vb \in \R^{m \times k}$& The matrix of  latent features for items\\
\hline
$\Sb \in \{-1, +1\}^{n \times n}$& The social network between $n$ users \\
\hline 
$\Omega_\Sb$, $|\Omega_\Sb|$ & The set of extracted triplets from the social relations  and its size \\
\hline
$\Wb \in \R_{+}^{n \times n}$& The pairwise similarity  matrix between users \\
\hline 
$\N(u) \subseteq [n]$& Neighbors of user $u$ in the social graph\\
\hline 
$\N_{+}(u) \subseteq [n]$& The set of trusted neighbors  by user $u$ in the social graph\\
\hline 
$\N_{-}(u) \subseteq [n]$& The set of distrusted neighbors  by user $u$ in the social graph\\
\hline 
$\D: \R^k \times \R^k \rightarrow \R_{+}$& The measurement function used to assess the similarly of latent features\\
\hline\end{tabular}}
\end{center}
\end{table}

\section{Matrix Factorization based Recommender  Systems}\label{sec:mf}
This section provides a formal definition of  collaborative filtering, the primary recommendation method we are concerned with in this paper, followed by solution methods for low-rank factorization that are proposed in the literature to address the problem. \\

\subsection{Matrix Factorization for Recommendation}
In collaborative filtering we assume that there is  a set of $n$ users
$\U = \{u_1, \cdots, u_n\}$ and a set of $m$ items $\I = \{i_1, \cdots, i_m\}$ where  each
user $u_i$ expresses opinions about a set of items. In this paper, we assume opinions are expressed through an explicit numeric rating (e.g., scale from one to five), but
other rating methods such as hyperlink clicks  are possible as well. We are mainly interested in recommending a set of items for an active user such that  the user has not rated these items before. To this end,  we are aimed at learning a model from the existing ratings, i.e., \textit{offline phase}, and then use the learned model to generate recommendations for active users, i.e., \textit{online phase}. The rating information is summarized in an $n \times m$ matrix $\Rb \in \R^{n \times m}, 1 \leq i \leq n, 1 \leq j \leq m$ where the rows correspond to the users and the columns correspond to the items and $(p,q)$th entry is the  rate given by user $u_p$ to the item $i_q$. We note that the rating matrix is partially observed and it is sparse in most cases.

An efficient and effective approach to recommender systems is to factorize the user-item rating matrix $\Rb$ by a multiplicative of $k$-rank matrices $\Rb \approx \Ub \Vb^{\top}$, where $\Ub \in \R^{n\times k}$ and $\Vb \in \R^{m\times k}$  utilize the factorized user-specific and item-specific  matrices, respectively, to make further missing data prediction.  The main intuition  behind a low-dimensional factor model is that there
is only a small number of factors influencing the preferences,
and that a user's preference vector is determined by how
each factor applies to that user. This low rank assumption makes it possible to effectively recover the missing entires  in the rating matrix from the observed entries.  We note that the celebrated Singular Value Decomposition (SVD) method to factorize the rating matrix $\Rb$ is not applicable here due to the fact that  the rating matrix  is partially available and we  are  only allowed to utilize the observed entries in factorization process. There are two basic formulations to solve this problem: these are optimization based (see e.g.,~\cite{0001,0099,64,koren2009matrix}) and probabilistic~\cite{mnih2007probabilistic}. In the following subsections, we first review the optimization based framework  for matrix factorization and then discuss how it can be extended to incorporate trust information.

\subsection{Optimization based Matrix Factorization}
Let $\Omega_\Rb $ be the set of observed ratings  in the user-item matrix $\Rb \in \R^{n\times m}$, i.e.,
\[ \Omega_\Rb = \{(i, j) \in [n] \times [m]: R_{ij}\;\; \text{has been observed} \},\]
where $n$ is the number of users and $m$ is the number of items to be rated. In optimization based matrix factorization, the goal is to learn the latent matrices $\Ub$ and $\Vb$ by solving the following optimization problem:
\begin{eqnarray}
\label{eqn:mf}
\min_{\Ub, \Vb} \left[\L(\Ub, \Vb) =  \frac{1}{2}  \sum_{(i,j) \in \Omega_\Rb}^{}{\left( R_{ij} - U_{i,:}^{\top}V_{j,:}\right)^2 } + \frac{\lambda_{U}}{2} \|\Ub\|_{\rm{F}} + \frac{\lambda_{V}}{2} \|\Vb\|_{\rm{F}} \right],
\end{eqnarray}
where $\|\cdot\|_{\rm{F}}$ is the Frobenius norm of a matrix, i.e,     $\|\mathbf{A}\|_{\rm{F}}=\sqrt{\sum_{i=1}^n\sum_{j=1}^m |A_{ij}|^2}$.  The optimization problem in~(\ref{eqn:mf}) constitutes of three terms: the first term aims to minimize the inconsistency between the observed entries and  their corresponding value obtained by the factorized matrices. The last two terms  regularize the latent matrices for users and items, respectively. The parameters $\lambda_U$ and $\lambda_V$ are regularization parameters that are introduced  to control the regularization of latent matrices $\Ub$ and $\Vb$, respectively.  We would like to emphasize  that the problem in~(\ref{eqn:mf}) is non-convex  jointly in both $\Ub$ and $\Vb$. However, despite its non-convexity, the formulation in~(\ref{eqn:mf}) is widely used in practical collaborative filtering applications as the performance is competitive or better as compared to trace-norm minimization, while scalability is much better. For example, as indicated in~\cite{koren2009matrix}, to address the Netflix problem, (\ref{eqn:mf}) has been applied with a fair amount of success to factorize data sets with 100 million ratings.
\subsection{Matrix Factorization with Trust Side Information} \label{sec:mftrust}
Recently it has been shown that just relying on the rating matrix to build a recommender system is not as accurate as expected. The main reason for this claim is the known cold-start users problem and the sparsity of rating matrix.  Cold-start users are one of the most important challenges in recommender systems. Since cold-start users are more dependent on the social network compared to users with more ratings, the effect of using trust propagation gets more important for cold-start users. Moreover, in many real life systems a very large portion of users do not express any ratings, and they only participate in the social network.  Hence, using only the observed ratings does not allow to learn the user features.

One of the most prominent  approaches to tackle the data sparsity problem in matrix factorization is to compensate the lack of information in rating matrix with other sources of side information which are available to the recommender system. It has been recently shown that social information such as trust relationship between users is a rich source of side information to compensate for the sparsity.  The above mentioned traditional recommendation techniques are all based on working on the user-item rating matrix, and ignore the abundant relationships among users.  
Trust-based recommendation usually involves constructing a trust network where
nodes are users and edges represent the trust placed on them. The goal of a trust-based
recommendation system is to generate personalized recommendations by aggregating
the opinions of other users in the trust network. The intuition is that users tend to adopt items recommended by trusted friends rather than strangers, and that trust is positively and strongly correlated with user preferences. Recommendation techniques that
analyze trust networks were found to provide very accurate and highly personalized
results. 

To incorporate the social relations in the optimization problem formulated in (\ref{eqn:mf}), few papers~\cite{44444,60,62,0099,zhu2011social} proposed the social regularization method which aims at keeping the latent vector of each user similar to his/her neighbors in the social network. The proposed models force the user feature vectors to be close to those of their neighbors to be able to learn the latent user features for users with no or very few ratings~\cite{60}. More specifically, the optimization problem becomes as:
\begin{align}
\label{eqn:social1}
\L(\Ub, \Vb) &= \frac{1}{2} \sum_{(i, j) \in \Omega_\Rb}^{}{\left( R_{ij} - U_{i,:}^{\top}V_{j,:}\right)^2 } + \frac{\lambda_{U}}{2} \|\Ub\|_{\rm{F}} + \frac{\lambda_{V}}{2} \|\Vb\|_{\rm{F}} \\
&\hspace{0.5cm} + \frac{\lambda_{S}}{2} \sum_{i = 1}^{n}{\left\| U_{i,:} - \frac{1}{|\N(i)|} \sum_{j \in \N(i)}^{}{U_{j,:}}\right\|}, \nonumber
\end{align}
where $\lambda_{S}$ is the social regularization parameter and $\N(i)$ is the subset of users who has relationship with $i$th  user in the social graph.

The rationale behind  social regularization idea is that   every user's taste is relatively similar to the average taste of his friends in the social network. We note that using this idea, latent features of users indirectly connected in the social network will be dependent and hence the trust gets propagated. A more reasonable and realistic model should treat all friends differently based on how similar they are. Let assume the weight of relationship between two users $i$ and $j$ is captured by $W_{ij}$ where $\mathbf{W} \in \R^{n \times n}$ demotes the social weight matrix. It is easy to extend the model in (\ref{eqn:social1}) to treat  friends differently based on the weight matrix $\mathbf{W}$ as:
\begin{align}
\label{eqn:social2}
\L(\Ub, \Vb) &= \frac{1}{2}  \sum_{(i, j) \in \Omega_\Rb}^{}{ \left( R_{ij} - U_{i,:}^{\top}V_{j,:}\right)^2 } + \frac{\lambda_{U}}{2} \|\Ub\|_{\rm{F}} + \frac{\lambda_{V}}{2} \|\Vb\|_{\rm{F}} \\
&\hspace{0.5cm} + \frac{\lambda_{S}}{2} \sum_{i = 1}^{n}{\left\| U_{i,:} - \frac{\sum_{j \in \N(i)}^{}{W_{ij}U_{j,:}}}{\sum_{j\in \N(i)}^{} W_{ij}} \right\|} \nonumber
\end{align}
An alternative formulation is to regularize each users' fiends individually, resulting in the following objective function~\cite{62}:
\begin{align*}
\L(\Ub,\Vb) &=  \frac{1}{2} \sum_{(i, j) \in \Omega_\Rb}^{}{\left( R_{ij} - U_{i,:}^{\top}V_{j,:}\right)^2 }  + \frac{\lambda_{U}}{2} \|\Ub\|_{\rm{F}} + \frac{\lambda_{V}}{2} \|\Vb\|_{\rm{F}}\\
&\hspace{0.5cm} + \frac{\lambda_S}{2} \sum_{i, j=1}^{n}{W_{ij}\left\| U_{i,:} - U_{j,:}\right\|^2}.
\end{align*}
where we simply assumed that for any $j \notin \N(i)$, $W_{ij} = 0$.

As mentioned earlier, the objective function in $\L(\Ub, \Vb)$ is not jointly convex in both $\Ub$ and $\Vb$ but it is convex in each of them fixing the other one. Therefore, to find a local solution one can  stick to  the standard gradient descent method to find a solution in an iterative manner as follows:
\begin{align*}
& \Ub_{t+1} \leftarrow \Ub_t - \eta_t \nabla_{\Ub} \L(\Ub, \Vb)\large{|}_{\Ub = \Ub_t, \Vb = \Vb_t}, \\
& \Vb_{t+1} \leftarrow \Vb_t - \eta_t \nabla_{\Vb} \L(\Ub, \Vb)\large{|}_{\Ub = \Ub_t, \Vb = \Vb_t}. \\
\end{align*}
\section{Matrix Factorization with Trust and Distrust Side Information}\label{sec:mftrust-distrust}
In this section we describe the proposed algorithm for social recommendation which is able to incorporate both trust and distrust relationships in the social network along with the partially observed rating matrix.   We then present two strategies to solve the derived optimization problem, one based on the gradient descent  optimization algorithm which generates more accurate solutions but it is computationally cumbersome,  and another based on the stochastic gradient descent  method which is computationally more efficient for large rating and social matrices but suffers from slow convergence rate.

\subsection{Algorithm Description}

As discussed before, the vast majority of related  work in the field of matrix factorization for recommendation has primarily focussed on trust propagation and simply ignore the distrust information between users, or intrinsically, are not capable of exploiting it.  Now, we aim at  developing a matrix factorization based model for recommendation in social rating networks to utilize both trust and distrust relationships. We incorporate the   trust/distrust relationship between users in our model to improve the quality of  recommendations. While intuition and experimental evidence indicate that trust is somewhat transitive, distrust is certainly not transitive. Thus, when we intend to propagate distrust through a network, questions about transitivity and how to deal with conflicting information abound.

To inject social influence in our model,  the basic idea is to find appropriate latent features  for users such that each user is brought closer to the users  she/he trusts  and separated apart from the users that she/he  distrusts and have different interests.  We note that simply incorporating this  idea in matrix factorization by naively penalizing the similarity of each user's latent features to his distrusted friends' latent features  fails to reach the desired goal.  The main reason is that   distrust is not as transitive as trust,  i.e. distrust can not directly replace trust in trust propagation approaches and utilizing distrust requires careful consideration (trust is transitive, i.e., if user $u$ trusts user $v$ and $v$ trusts $w$, there is a good chance that $u$ will trust $w$, but distrust is certainly not transitive, i.e., if $u$ distrusts $v$ and $v$ distrusts $w$, then $w$ may be closer to $u$ than $v$ or maybe even farther away).  It is noticeable that this statement is consistent with  the preliminary experimental results in~\cite{VictorCCT11} for memory-based CF methods that indicate  regarding distrust as an indication to reverse deviations in not the right way to incorporate distrust. Therefore we pursue another approach to model the distrust in recommendation process.

The main intuition behind the proposed framework stems from the observation that the trust relations between users can be treated as {\it{agreement}} on items and distrust  relations can be considered  as {\it{disagreement}} on items. Then, the question becomes how can we guarantee  when a user agrees on an item with one of his/her friends, he/she will disagree on the same item with his/her distrusted friends with a reasonable margin. We note that this margin   should be large enough to make it possible to distinguish between two types of friends.  In terms of latent features, this observation translates to having a margin between the similarity and dissimilarity of users' latent features to his/her trusted and distrusted friends.

Alternatively,  one can view the proposed method from the viewpoint of connectivity of latent features in a properly designated graph. Intuitively, certain features or groups of features should influence how users connect in the social network, and thus it should be possible to learn a mapping from features to connectivity in the social network such that the mapping respects the underlying  structure of the social network. In the basic matrix factorization algorithm for recommendation, we can consider the latent features as isolated vertices of a graph where there is no connection between nodes. This can be generalized  to the social-enhanced setting by  considering the social graph as the underlying graph between latent features with two types of edges (i.e., trust and distrust relations correspond to positive and negative edges, respectively).  Now the problem reduces  to learning the latent features for each user $u$ such that users trusted by $u$ in the social network (with positive edges) are close and users which are distrusted by $u$ (with negative edges) are more distant. Learning latent features in this manner  respects   the inherent topology of the social network. 

\begin{figure}[t]
\centering
\subfigure[User trust netwrok]{\includegraphics[scale=0.3]{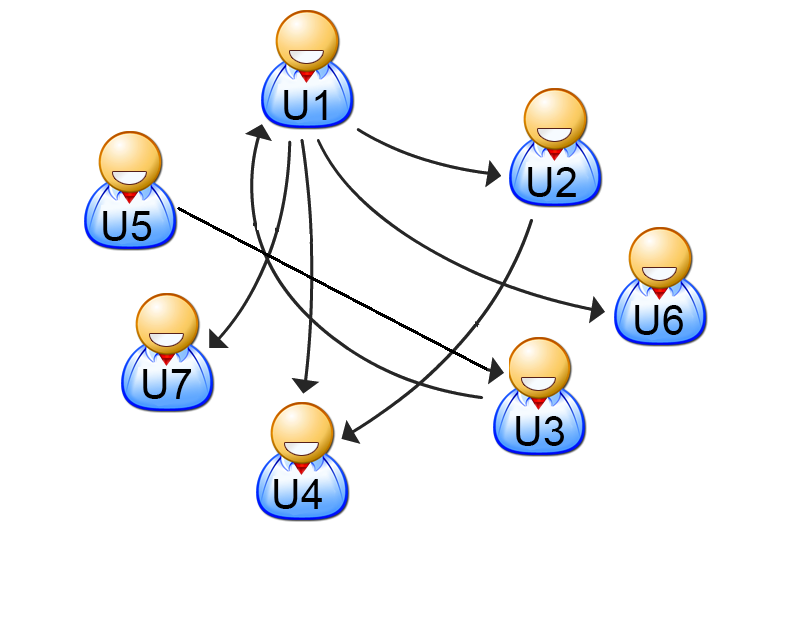}}
\subfigure[User distrust netwrok]{\includegraphics[scale=0.28]{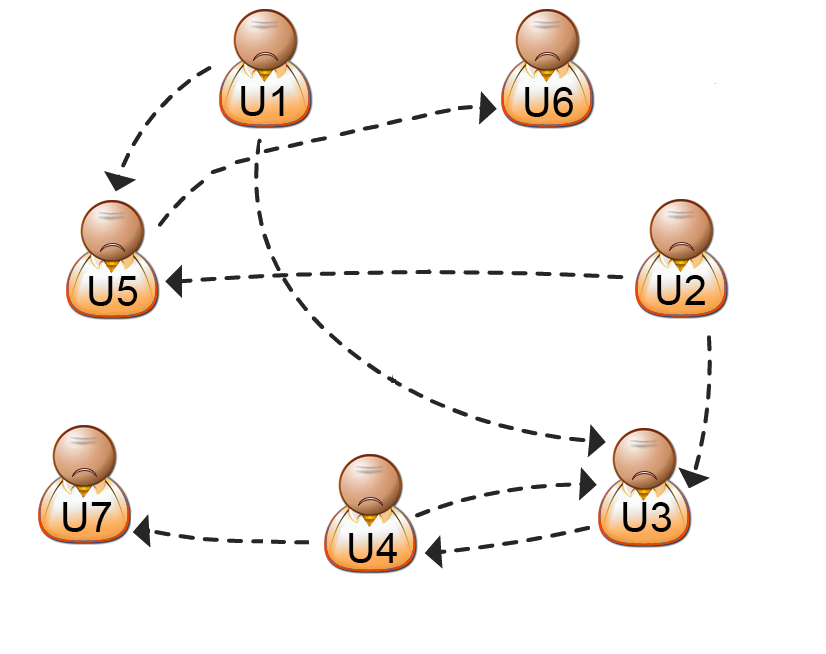}}
\subfigure[Partially observed rating matrix]{\includegraphics[scale=0.15]{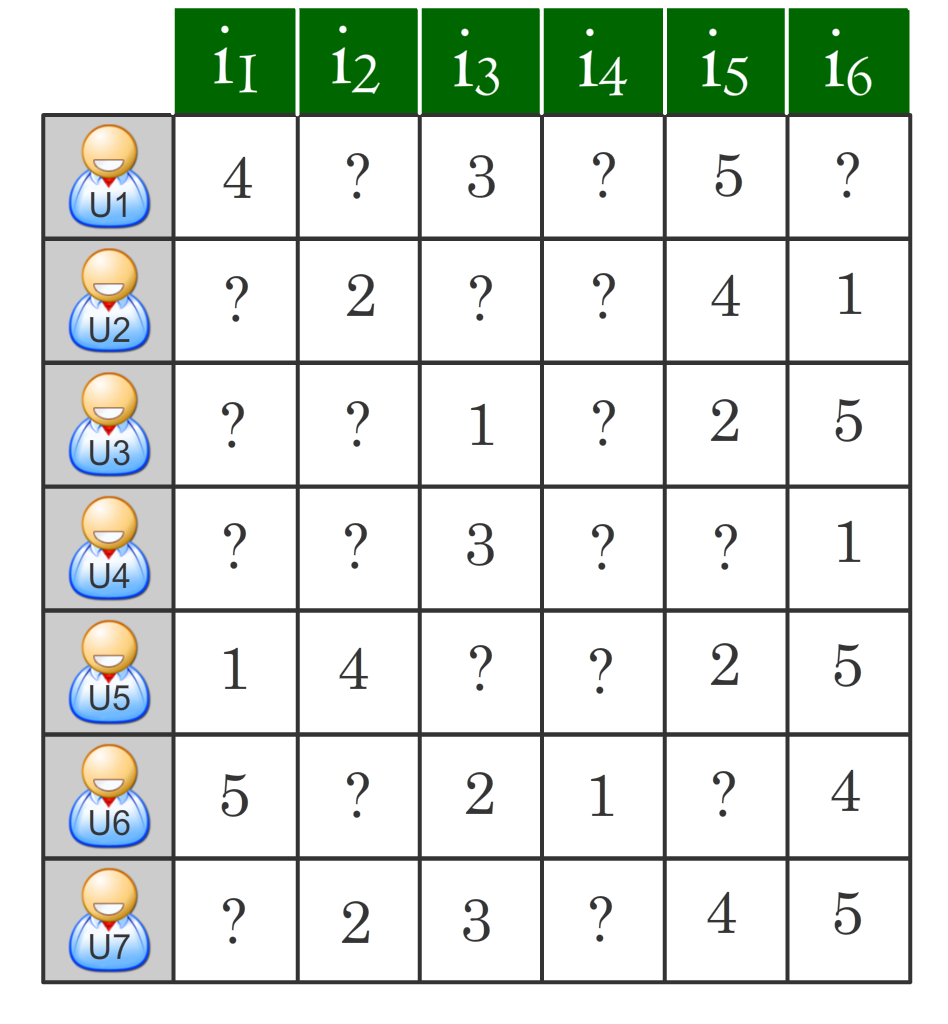}} \hspace{0.5cm}
\subfigure[Illustration of learned latent features]{\includegraphics[scale=0.13]{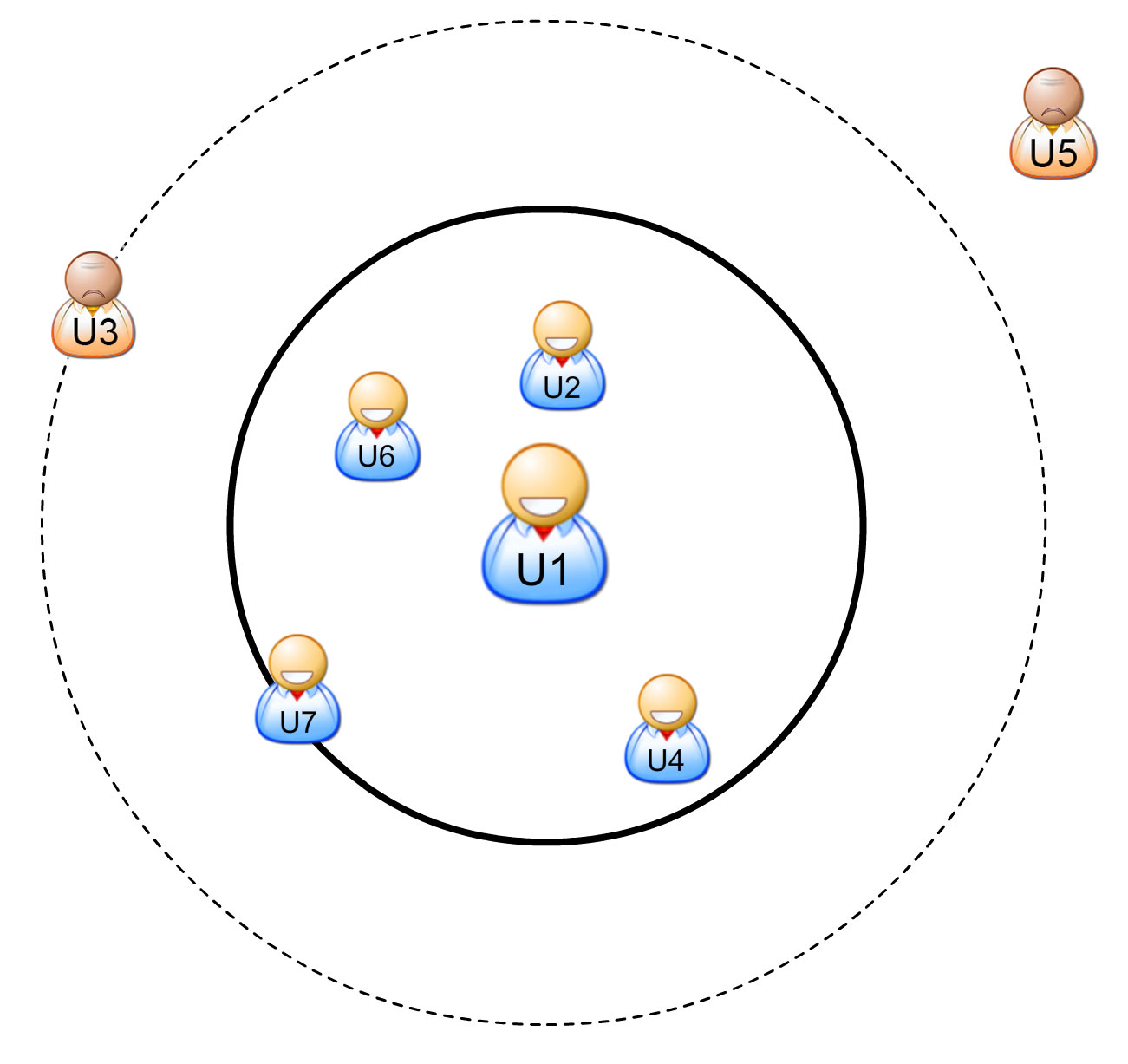}}

\caption{A simple example with seven users $\{u_1, u_2, \cdots,u_7\}$ and  six items $\{i_1, i_2, \cdots,i_6\}$ to illustrate the main intuition behind the proposed algorithm. The inputs of the algorithm are (a) trust network, (b) distrust network, and (c) partially observed rating matrix $\Rb$, respectively. As shown in (d) for user $u_1$ , the learned latent features for all his trusted friends $\{u_2,u_4,u_6,u_7\}$ are closer to $u_1$'s latent features than his distrusted friends $\{u_3,u_5\}$  with a margin of 1.}
\label{fig:sample}
\end{figure}

Figure~\ref{fig:sample} shows an example to illustrate the intuition behind the mentioned idea.  For  ease of exposition, we only consider the latent features  for the user $u_1$. From the trust network in Figure~\ref{fig:sample} (a)  we can see that  user $u_1$ trusts the list of users $\N_{+} = \{u_2, u_4, u_6, u_7\}$ and from the distrust network in Figure~\ref{fig:sample} (b) we  see that user $u_1$ distrusts the list of users $\N_{-} = \{u_3, u_5\}$. The goal is to learn the latent features that obeys two goals, i) it minimizes the prediction error on observed entries in the rating matrix, ii) it respects the underlying  structure of the trust and distrust networks between users. In Figure~\ref{fig:sample} (d) the latent features are depicted in the Euclidean space from the viewpoint of user $u_1$.  As shown in  Figure~\ref{fig:sample} (d), for user $u_1$, the latent features of his/her trusted friends $\N_{+}$ lie inside the solid circle centered at $u_1$ and the latent features of his/her distrusted friends $\N_{-}$ lie outside the dashed circle.  The gap between two circles guarantees that always there exists a safe margin between $u_1$'s agreements with his trusted and distrusted friends. One simple way to impose  these constraints on the latent features of users is to generate a set of triplets  for any combination of trusted  and distrusted friends ( e.g., one such triplet  for  user  $u_1$  can be constructed as $(u_1, u_2, u_5)$) and force the margin constraint to hold for all extracted triplets. This ensures that the minimum margin gap will definitely exist between the latent features of all the trusted and distrusted friends as desired and makes it possible to incorporate both types of relationships between users  in the matrix factorization.

It is worthy to mention that similar to the  social-enhanced recommender systems discussed before, the proposed  algorithm is also based on hypotheses about the existence and the correlation of trust/distrust relations and ratings in the data.  The empirical investigation of correlation between social relations and rating information has been the focus of a bulk of recent research including~\cite{ziegler2007investigating,patil2013quantifying,ma2013experimental}, where the results   reinforce the hypothesis that  ratings from trusted people count more than those from others and in particular distrusted neighbors. We have also conducted experiments as will be detailed in Subsection~\ref{sec:emp-cons}, to empirically investigate the correlation/alignment between social relations and the rating information issued by users which supports our strategy in exploiting the trust/distrust relations in matrix factorization.

We now formalize the proposed solution.  As the first ingredient, we need a measure to evaluate the consistency between the latent features of users, i.e., the matrix $\Ub$, and the trust and distrust constraints   existing between  users in the social network. To this end,  we introduce a monotonically increasing convex loss function $\ell(z)$ to measure the discrepancy between the latent features of different users.  Let $u_i$, $u_j$, and $u_k$ be three users  in the model such that $u_i$ trusts $u_j$ but distrusts $u_k$. The main intuition behind the proposed framework is that the latent features of $u_i$, i.e., $U_{i,:}$ must be more similar to $u_j$'s latent features than latent features for user $u_k$.  For each such a triplet we penalize the objective function by $\ell(\D(U_{i,:}, U_{j,:}) - \D(U_{i,:}, U_{k,:}))$ where the function  $\D: \R^k \times \R^k \mapsto \R_{+}$ measures the similarity between two latent vectors assigned to two different users, and $\ell:\R \mapsto \R_{+}$ is a penalty function that is  utilized to assess the violation of latent vectors of trusted and distrusted users. Example loss functions include hinge loss $\ell(z) = \max (0, 1 - z)$ and logistic loss $\ell(z) = \log (1 + e^{-z})$ which are widely used convex surrogate of 0-1 loss function in  learning community.

Let  $\Omega_{\Sb}$ denote the set of extracted triplets from the social relations, i.e., 
\[ \Omega_{\Sb} = \left\{ (i, j ,k) \in [n]\times[n]\times[n]: S_{ij} = 1 \;\;\&\;\; S_{ik} = -1 \right\}.\]
Here, a positive relationship means friends or a trusted relationship and  a negative relationship means foes or a distrust relationship. Then, our goal becomes to find a factorization of matrix $\Rb$ such that  the learned latent features of users are consistent with the constraints in $\Omega_{\Sb}$ where the consistency is reflected in the loss function. This results in the following optimization problem:
\begin{align}
\label{eqn:general}
\L(\Ub,\Vb) &=  \frac{1}{2} \sum_{(i, j) \in \Omega_{\Rb}}^{}{\left( R_{ij} - U_{i,:}^{\top}V_{j,:}\right)^2 }  + \frac{\lambda_{U}}{2} \|\Ub\|_{\rm{F}} + \frac{\lambda_{V}}{2} \|\Vb\|_{\rm{F}} \nonumber \\
&\hspace{0.5cm} + \frac{\lambda_S}{|\Omega_{\Sb}|} \sum_{(i,j,k) \in \Omega_{\Sb}}^{}{\ell(\D(U_{i,:}, U_{j,:}) - \D(U_{i,:},U_{k,:}))}.
\end{align}
Let us make the above general formulation more  specific by setting $\ell(\cdot)$ and $\D(\cdot, \cdot)$ to be the hinge loss and  the Euclidian distance, respectively.  Under these two assumptions, the  objective can be formulated as:
\begin{align}
\label{eqn:hing-euc}
\L(\Ub,\Vb) &=  \underbrace{\frac{1}{2} \sum_{(i, j) \in \Omega_{\Rb}}^{}{\left( R_{ij} - U_{i,:}^{\top}V_{j,:}\right)^2 }}_{\mathcal{R}(\Ub,\Vb)}  + \frac{\lambda_{U}}{2} \|\Ub\|_{\rm{F}} + \frac{\lambda_{V}}{2} \|\Vb\|_{\rm{F}} \nonumber\\
&\hspace{0.5cm} + \frac{\lambda_S}{|\Omega_{\Sb}|} \sum_{(i,j,k) \in \Omega_{\Sb}}^{}{\max\left(0, 1 - \|U_{i,:}-U_{j,:}\|^2 + \|U_{i,:} - U_{k,:} \|^2\right)}.
\end{align}
Here the constraints have been written in terms of hinge-losses over triplets, each consisting of a user, his/her trusted friend  and his/her distrusted friend. Solving the optimization problem in~(\ref{eqn:hing-euc})  outputs the latent features for users and items that can utilized to estimate the missing values in the user-item matrix. Comparing the formulation in~(\ref{eqn:hing-euc}) to the existing factorization-based methods discussed earlier reveals two main features of the proposed formulation. First, it aims to minimize the error on the observed ratings and to respect the inherent structure of the social network among the users. The trade-off between these two objectives is captured by the regularization parameter $\lambda_S$ which is required to be tuned effectively.

In a similar way,   applying the logistic loss to the general formulation in~(\ref{eqn:general}) yields the following objective:
\begin{align}
\label{eqn:logit-loss}
\L(\Ub,\Vb) &=  \frac{1}{2} \sum_{(i, j) \in \Omega_{\Rb}}^{}{\left( R_{ij} - U_{i,:}^{\top}V_{j,:}\right)^2 }  + \frac{\lambda_{U}}{2} \|\Ub\|_{\rm{F}} + \frac{\lambda_{V}}{2} \|\Vb\|_{\rm{F}} \nonumber\\
&\hspace{0.5cm} + \frac{\lambda_S}{|\Omega_{\Sb}|} \sum_{(i,j,k) \in \Omega_\Sb}^{}{\log \left( 1+ \exp \left(  \|U_{i,:} - U_{k,:} \|^2 - \|U_{i,:}-U_{j,:})\|^2 \right) \right)}.
\end{align}
\begin{remark} We note that in several applications of recommender systems, besides the observed ratings, a description of the users and/or the objects through attributes (e.g., gender, age) or measures of similarity is available that could potentially benefit the process of recommendation (see e.g.~\cite{agarwal2010flda} for few interesting applications). In that case it is tempting to take advantage of both known ratings and descriptions to model the preferences of users.    A natural way to incorporate the available meta-data is to kernalize the similarity measure between latent features based on a positive definite kernel between pairs that can be deduced from the meta-data. More specifically,  instead of simply using  Euclidian distance as the similarity measure between latent features in (\ref{eqn:hing-euc}), we can use the kernel matrix $\mathbf{K}$ obtained from the  Laplacian of the  graph obtained from the meta-data  to measure the similarity as:
\[ \D(U_{i,:}, U_{j,:}) = \left(U_{i,:} - U_{j,:}\right)^{\top} \mathbf{K} \left(U_{i,:} - U_{j,:}\right),\]
where $\mathbf{K} = (\mathbf{D} - \mathbf{W})^{-1}$, with $\mathbf{D}$ as a diagonal matrix with $D_{i,i} = \sum_{j=1}^{n}{W_{ij}}$. Here $\mathbf{W}$ captures the pairwise weight between users  in the similarity graph between  users that is computed based on the available meta-data about users. 
\end{remark}
\begin{algorithm}[t]
\center \caption{GD based Matrix Factorization with Trust and Distrust Propagation}
\begin{algorithmic}[1] \label{alg:1}
    \STATE {\bf Input}: $\Rb$: partially observed rating matrix, $\Omega_{\Sb}$
    \STATE {\bf Output}: $\Ub$ and $\Vb$
    \FOR{$t = 1, \ldots, T$}	 
	\STATE Compute the gradients  $\nabla_{\Ub}\mathcal{R}(\Ub_t, \Vb_t)$ and $\nabla_{\Vb}\mathcal{R}(\Ub_t, \Vb_t)$.
	\STATE Compute $\nabla_{\Ub}$ by Eq.~\ref{eqn:gradU}
	\STATE Compute $\nabla_{\Vb}$ by Eq.~\ref{eqn:gradV}
         \STATE Update:
        \[\Ub_{t+1} = \Ub_t - \eta_t \nabla_{\Ub} |_{\Ub = \Ub_t, \Vb = \Vb_t}\]  
         \[\Vb_{t+1} = \Vb_t - \eta_t \nabla_{\Vb} |_{\Ub = \Ub_t, \Vb = \Vb_t}\]
    \ENDFOR
    \STATE {\bf return}  $\Ub_{T+1}$ and $\Vb_{T+1}$.
\end{algorithmic}
\end{algorithm}
\begin{remark} We would like to emphasize  that it is straightforward to generalize the proposed framework to incorporate  similarity and dissimilarity information between items. What we need is to extract the triplets from the trust/distrust links between items and repeat the same process  we did for users. This will add another term to the  objective in terms of latent features of items $\Vb$ as shown in the following generalized formulation:
\begin{align*}
\L(\Ub,\Vb) &=  {\frac{1}{2} \sum_{(i, j) \in \Omega_{\Rb}}^{}{\left( R_{ij} - U_{i,:}^{\top}V_{j,:}\right)^2 }} + \frac{\lambda_{U}}{2} \|\Ub\|_{\rm{F}} + \frac{\lambda_{V}}{2} \|\Vb\|_{\rm{F}} \\
&\hspace{0.5cm} + \frac{\lambda_S}{|\Omega_{\Sb}|} \sum_{(i,j,k) \in \Omega_{\Sb}}^{}{\max\left(0, 1 - \|U_{i,:}-U_{j,:}\|^2 + \|U_{i,:} - U_{k,:} \|^2\right)} \\
&\hspace{0.5cm} + \frac{\lambda_{{I}}}{|\Omega_{\mathbf{I}}|} \sum_{(i,j,k) \in \Omega_{\mathbf{I}}}^{}{\max\left(0, 1 - \|V_{i,:}-V_{j,:}\|^2 + \|V_{i,:} - V_{k,:} \|^2\right)},
\end{align*}
where $\lambda_{I}$ is the regularization parameter and $\Omega_{\mathbf{I}}$ is the set of triplets extracted from the similar/dissimilar links between items. The similarity/dissimilarity   links between items can be constructed according to   tags issued by users or associated with items, and categories. For example, if two items are attached with a same tag, there is a trust link between them and otherwise distrust link. Alternatively, trust/distrust links can be extracted  by measuring  similarity/dissimilarity based on the item properties or profile if provided.  This can further improve the accuracy of recommendations. 
\end{remark}

\subsection{Batch Gradient Descent based Optimization}
In optimization for supervised machine learning, there exist two regimes in which popular
algorithms tend to operate: the stochastic approximation regime, which samples
a small data set per iteration, typically a single data point, and the batch or sample
average approximation regime, in which larger samples are used to compute an
approximate gradient. The choice between these two extremes outlines the well-known
tradeoff between inexpensive noisy steps and expensive but more reliable steps. Two preliminary examples of these regimes are the Gradient Descent (GD) and the Stochastic Gradient Descent (SGD)   methods, respectively.  Both GD and SGD methods starts with some initial point, and iteratively updates the solution using the gradient information at intermediate solutions. The main difference is that GD requires a full gradient information at each iteration while SGD only requires an unbiased estimate of the full gradient which can be done by sampling

We now  discuss the application of  GD algorithm to solve the optimization problem in~(\ref{eqn:hing-euc}) as detailed in Algorithm~\ref{alg:1}. Recall that the objective function is not jointly convex in both $\Ub$ and $\Vb$. On the other hand, the objective is convex in one parameter by fixing the other one. Therefore, we follow an iterative method to minimize the objective. At each iteration, first by fixing  $\Vb$, we take a step in the direction of the negative gradient for $\Ub$ and repeat the same process for $\Vb$ by fixing  $\Ub$. 

For the ease of exposition, we introduce further notation.  For any triplet $(i,j,k) \in \Omega_\Sb$ we note that the $\|U_{i,:} - U_{j,:} \|^2 - \|U_{i,:}-U_{k,:}\|^2$ can be written as $\tr(\Cb \Ub^{\top} \Ub)$ where $\tr(\cdot)$ denotes the trace of the input matrix and $\Cb$ is a sparse auxiliary matrix defined for each triplet with all entries equal to zero except: $\Cb_{ik} = \Cb_{ki} = \Cb_{jj} = 1$ and $\Cb_{kk} = \Cb_{ij} = \Cb_{ji} = -1$. Having defined this notation, we can write the objective in~(\ref{eqn:hing-euc}) as:

\begin{align*}
\L(\Ub,\Vb) &=  \mathcal{R}(\Ub,\Vb)  + \frac{\lambda_{U}}{2} \|\Ub\|_{\rm{F}} + \frac{\lambda_{V}}{2} \|\Vb\|_{\rm{F}} + \frac{\lambda_S}{|\Omega_{\Sb}|} \sum_{(i,j,k) \in \Omega_{\Sb}}^{}{\max\left(0, 1 - \tr(\Cb_{ij}^k \Ub^{\top} \Ub) \right)}.
\end{align*}
where $\Cb_{ij}^k$ is the $\Cb$ matrix defined above which is associated with triplet $(i,j,k)$.  To apply the GD method, we need to compute the gradient of  $\L(\Ub,\Vb) $ with respect to $\Ub$ and $\Vb$ which we denote by $\nabla_{\Ub} = \nabla_{\Ub} \L(\Ub, \Vb)$ and $\nabla_{\Vb} = \nabla_{\Vb} \L(\Ub, \Vb)$, respectively.  We have:

\begin{align}\label{eqn:gradU}
\nabla_{\Ub} =\nabla_{\Ub} \mathcal{R}(\Ub,\Vb)+  \lambda_U \Ub - \frac{\lambda_S}{|\Omega_{\Sb}|} \sum_{(i,j,k) \in \Omega_{\Sb}}^{}{\mathbf{1}_{[\tr(\Cb_{ij}^k \Ub^{\top} \Ub) < 1]} (\Ub \Cb_{ij}^{k{\top}} + \Ub \Cb_{ij}^k)}
\end{align}
where $\mathbf{1}_{[\cdot]}$ is the indicator function which takes a value of one if its
argument is true, and zero otherwise. Similarly for $\nabla_{\Vb}$ we have:
\begin{align}\label{eqn:gradV}
\nabla_{\Vb} =\nabla_{\Vb} \mathcal{R}(\Ub,\Vb)+  \lambda_V \Vb 
\end{align}
The main shortcoming of GD  method is its high computational cost per iteration due to the gradient computation (i.e., step~(\ref{eqn:gradU})) which is  expensive when the  size of social constraints $\Omega_{\Sb}$ is large. We note that the size of $\Omega_{\Sb}$ can be as large as $O(n^3)$ by considering all triplets in the social graph. In the next subsection we provide an alternative solution to resolve this issue  using the stochastic gradient descent  and mini-batch SGD methods which are more efficient than the GD method in terms of the computational cost per iteration but with a slow convergence rate in terms of target approximation error. 
\begin{algorithm}[t]
\center \caption{Mini-SGD based Matrix Factorization with Trust and Distrust Propagation}
\begin{algorithmic}[1] \label{alg:2}
    \STATE {\bf Input}: $\Rb$: partially observed rating matrix, $\Omega_{\Sb}$, min batch size $B$
    \STATE {\bf Output}: $\Ub$ and $\Vb$

    \FOR{$t = 1, \ldots, T$}
    	   \STATE $\nabla_t \leftarrow \mathbf{0}$
	    \FOR{$b = 1, \ldots, B$}
	    	\STATE $(i, j, k) \leftarrow\;\text{Sample random triplet from}\;\; \Omega_{\Sb}$
		\IF {$(1 - \|U_{i,:}-U_{j,:})\|^2 + \|U_{i,:} - U_{k,:} \|^2 > 0)$}
			\STATE $\nabla_t \leftarrow \Ub_t \Cb_{ij}^k \Ub_t^{\top} $
		\ENDIF
  	    \ENDFOR
	\STATE Compute the gradients  $\nabla_{\Ub}\mathcal{R}(\Ub_t, \Vb_t)$ and $\nabla_{\Vb}\mathcal{R}(\Ub_t, \Vb_t)$.
        \STATE Update:
        \[\Ub_{t+1} = \Ub_t - \eta_t \left( \nabla_{\Ub}\mathcal{R}(\Ub_t, \Vb_t) +\lambda_U \Ub_t + \frac{\lambda_S}{B|\Omega_{\Sb}|}\nabla_t\right)\]  
        \STATE Update: 
         \[\Vb_{t+1} = \Vb_t - \eta_t \left(\nabla_{\Vb}\mathcal{R}(\Ub_t, \Vb_t) +\lambda_V \Vb_t  \right)\]
    \ENDFOR
    \STATE {\bf return}  $\Ub_{T+1}$ and $\Vb_{T+1}$.
\end{algorithmic}
\end{algorithm}
\subsection{Stochastic and Mini-batch Optimization}
As discussed above, when the size of social network is very large, the size of $\Omega_{\Sb}$  may cause computational  problems in solving the optimization problem in~(\ref{eqn:hing-euc}) using GD method. The reason is essentially the fact that computing the gradient at each iteration requires to go through all the triplets in $\Omega_{\Sb}$ which is infeasible for large  networks. To alleviate this problem we propose a stochastic gradient based~\cite{nemirovski-2009} method to solve the optimization problem. The main idea is to choose a fixed subset of triplets for gradient computation instead of all $|\Omega_{\Sb}|$  triplets at \textit{each} iteration~\cite{cotter2011better}. More specifically, at each iteration, we sample $B$ triplets uniformly at random from $\Omega_{\Sb}$ to compute the next solution. We note that this strategy generates unbiased estimates of the true gradient and makes each iteration of algorithm computationally more efficient compared to the full gradient counterpart. In the simplest case, SGD  algorithm, only one triplet is chosen at each iteration to generate an unbiased estimate of the full gradient.  We note that in practice SGD is usually implemented based on data shuffling, i.e., making the sequence of the training samples random and then training the model by going through the training samples one by one.  An intermediate solution, known as mini-batch SGD, chooses a subset of triplets  to compute the gradient. The promise is that by selecting more triplets at each iteration, on one hand the variance of stochastic gradients  decreases promotional to the number of sampled triplets, and on the other hand the algorithm enjoys the light computational cost of basic SGD method.    

The detailed steps of the algorithm are shown in Algorithm~\ref{alg:2}. The mini-batch SGD method improves the computational efficiency  by grouping multiple constraints into a mini-batch and only updating the $\Ub$ and $\Vb$ once for each mini-batch. For brevity, we will refer to this algorithm as Mini-SGD.  More specifically, the Mini-SGD algorithm, instead of computing the full gradient over all triplets,  samples $B$ triplets uniformly at random from $\Omega_{\Sb}$ where $1 \leq B \leq |\Omega_{\Sb}|$ is a parameter that needs to be provided to the algorithm, and computes the stochastic gradient as:

\[ \nabla_t =  \frac{\lambda_S}{B} \sum_{(i,j,k) \in \Omega_B}^{}{\mathbf{1}_{[\tr(\Cb_{ij}^k \Ub_t^{\top} \Ub_t) <  1]} (\Ub \Cb_{ij}^{k{\top}} + \Ub \Cb_{ij}^k)}\]

where $\Omega_B$ is the set of $B$ sampled triplets from $\Omega_{\Sb}$. We note that 
\[ \mathbf{E} [\nabla_t] = \frac{\lambda_S}{|\Omega_{\Sb}|} \sum_{(i,j,k) \in \Omega_{\Sb}}^{}{\mathbf{1}_{[\tr(\Cb_{ij}^k \Ub_t^{\top} \Ub_t)  < 1]} (\Ub \Cb_{ij}^{k{\top}} + \Ub \Cb_{ij}^k)},\]
i.e., $\nabla_t$ is an unbiased estimate of the full gradient in the right hand side. When $B = |\Omega_{\Sb}|$, each iteration handles the original objective function and Mini-SGD reduces to the batch GD algorithm. We note that both GD and SGD share the same convergence rate in terms of number of iterations in expectation for non-smooth optimization problems (i.e., $O(1/\sqrt{T})$ after $T$ iterations), but SGD method requires much less running time to convergence  compared to the GD method  due to the efficiency of its individual iterations.

\section{Experimental Results}\label{sec:exp}
In this section, we conduct exhaustive experiments to demonstrate the merits and advantages of the proposed algorithm. We conduct the experiments on  the well-known  Epinions~\footnote{\url{http://www.trustlet.org/wiki/Epinions_datasets}} data set, aiming to accomplish and answer the following fundamental  questions:
\begin{enumerate}
\item \textbf{Prediction accuracy:} How does the proposed algorithm perform in comparison to the state-of- the-art algorithms with/without incorporating  trust and distrust relationships between users.  Whether or not the trust/distrust social network could help in making   more accurate recommendations? 
\item \textbf{Correlation of social relations with rating information:} To what extent, the trusted and distrusted friends of a  user $u$ are aligned with the ratings the user $u$ issued for the reviews written by his friends? A positive answer to this question  indicates that two users will issue similar (dissimilar) ratings if they are connected by a trust (distrust) relation and prefer to behave similarly. 

\item \textbf{Model selection:} What role do the regularization  parameters  $\lambda_S$, $\lambda_U$ and $\lambda_V$ play in the accuracy of the proposed recommender system and what is the best strategy to tune these parameters?

\item \textbf{Handling cold-start users:} How does exploiting social relationships in prediction process affect the performance of recommendation for cold-start users?

\item \textbf{Trading trust for distrust:} To what extent the distrust relations can compensate for the lack of trust relations?

\item \textbf{Efficiency  of optimization:} What is the trade-off between accuracy and efficiency by moving from the gradient descent to the stochastic gradient descent   with different batch sizes?

\end{enumerate}
In the following subsections, we intend to answer these questions.   We begin by introducing the data set we use in our experiemnts and the metrics we employ to evaluate the results, followed by the detailed experimental results.

\subsection{Data Set Description and Experimental Setup}

\paragraph{The Epinions data set} We begin by discussing the data set we have chosen for our experiments. To evaluate the proposed algorithm on trust and distrust-aware recommendations, we use the   Epinions data set~\cite{Guha-2004-propagation}, a popular e-commerce site and customer review website where users share opinions on various types of items such as electronic products, companies, and movies, through writing reviews about them or assigning a rating to  the reviews written by other users. The rating values in Epinions are discrete values ranging from Ònot helpfulÓ (1/5) to Òmost helpfulÓ (5/5). These ratings and reviews would potentially  influence future customers when they are about to decide whether a product is worth buying or a movie is worth watching.

Epinions allows users to evaluate other users based on the quality of their reviews, and to make trust and distrust relations with other users in addition to the ratings.  Every member of Epinions can maintain a "trust" list of people he/she trusts that is referred to as \textit{web of trust} (social network with trust relationships) based on the  reviewers with  consistent ratings or "distrust" list known as \textit{block list} (social network with distrust relationships) that presents reviewers whose reviews were consistently found to be inaccurate or low quality.  The fact that the data set contains explicit positive and negative relations between users  makes it very appropriate to study issues in trust- and distrust-enhanced recommender systems. Epinions is thus an ideal source for experiments on social recommendation.  We remark that the Epinions data set only contains bivalent relations (i.e., contains only full trust and full distrust,
and no gradual statements).

To conduct the coming experiments, we sampled a subset of Epinions data set with $n=121,240$ users and $m = 685,621$ different items. The total number of observed ratings in the sampled data set  is 12,721,437 which approximately includes $0.02\%$ of all entries in the rating matrix $\Rb$ which  demonstrates  the sparsity of the rating matrix.  We note that the selected items are the most frequently rated overall. The statistics of the data set is given in Table~\ref{table:2}.  The social statistics of the this data source is summarized in Table~\ref{table:4}. The frequencies  of ratings for users is shown are Table~\ref{table:stat}. In the user distrust network, the total number of issued distrust statements is 96,823.  As to the user trust network, the total number of issued trust statements is 481,799. 

\paragraph{Experimental setup}{To better evaluate the effect of utilizing the social side information in recommendation accuracy, we employ different amount of training data 90\%, 80\% , 70\% and 60\%   to create four different training sets that are increasingly sparse but the social network remains the same in all of them. Training data 90\%, for example, means we randomly select 90\% of the ratings from the sampled Epinions data set as the training data to predict the remaining 10\% of ratings. The random selection was carried out $5$ times independently to have a fair comparison.   Also, since our preliminary results on a smaller data set revealed that the hinge loss performs better than the exponential loss, in the rest of experiments we stick to this loss function. However, we note the exponential loss is slightly faster in optimizing the corresponding objective function  thanks to its smoothness, but it was negligible considering  its  worse accuracy compared to the hinge loss. All implementations are in Matlab, and all experiments were performed on a  4-core 2.0 GHZ of a load-free  machine with a 12G of RAM.}

\begin{table*}[t]
\begin{center}
\caption{Statistics of sample data from Epinions data set used in our experiments.\label{table:2}}{%
\begin{tabular} {| l || l |  }
\hline 
Statistic & Quantity \\ 
\hline 
Number of users & 121,240 \\ 
\hline
Number of items & 685,621 \\
\hline 
Number of ratings & 12,721,437 \\
\hline
Number of trust relations &  481,799 \\
\hline 
Number of distrust relations & 96,823 \\
\hline
Minimum number of ratings by users   & 1 \\
\hline 
Minimum number of ratings for items  & 1 \\
\hline
Maximum number of ratings  by  users & 148735 \\
\hline 
Maximum number of ratings for items &  945 \\
\hline
Average number of ratings by users & 85.08  \\
\hline 
Average number of ratings for items &  15.26 \\
\hline\end{tabular}}
\end{center}
\end{table*}
\begin{table*}[t]
\begin{center}
\caption{Maximum and average trust and distrust relations for users in the sampled data set.\label{table:4}}{%
\begin{tabular} {|l|c|c|  }
\hline  
Statistics & Trust per user & Be Trusted  per user \\
 \hline \hline 
Max & 1983 & 2941 \\
\hline 
Min & 1 & 0 \\ \hline 
Average &  4.76 & 4.76 \\
\hline \hline  
 & Distrust per user & Be Distrusted  per user \\
  \hline 
Max & 1188 & 429 \\
\hline 
Min & 1 & 0 \\ \hline 
Average & 0.91 & 0.91 \\
\hline
\end{tabular}}
\end{center}
\end{table*}

\begin{table*}[t]
\begin{center}
\caption{\ct{The frequencies of user's rating}.\label{table:stat}}{%
\begin{tabular} {|l|c|c|c|c|c|c|c|}
\hline  
\# of Ratings & 0-10 & 11-20 & 21-30 & 31-40 & 41-50 &  $ 51-60 $  \\ \hline 
\# of Users & 4,198,074 ($\approx 33\%$) & 3,053,144 ($\approx 24\%$)  & 2,289,858 ($\approx 18\%$)& 1,526,572 ($\approx 12\%$) & 534,300 ($\approx 4.2\%$) &  267,150 ($\approx 2.1\%$)    \\ \hline \hline 

\# of Ratings & 61-70 & 71-80 & 81-90 & 91-100 & 101-200 &  $  201-300$  \\ \hline 
\# of Users &  157,745 ($\approx 1.24\%$) &  143,752 ($\approx 1.13\%$)  &  104,315 ($\approx 0.82\%$)&  43,252 ($\approx 0.34\%$) &  21,626 ($\approx 0.17\%$) &  10,686 ($\approx 0.084\%$)    \\ \hline  

\hline
\end{tabular}}
\end{center}
\end{table*}

\subsection{Metrics}
\subsubsection{Metrics for rating prediction}
We employ two well-known measures, the Mean Absolute Error (MAE) and the Root Mean Squared Error (RMSE)~\cite{herlocker2004evaluating} to measure the prediction accuracy of the proposed approach in comparison with other basic collaborative filtering and trust/distrust-enhanced recommendation methods.

MAE is very appropriate and useful measure for evaluating prediction accuracy in offline tests~\cite{herlocker2004evaluating,20004}. To calculate MAE, the predicted rating is compared with the real rating and the difference (in absolute value) considered as the prediction error. Then, these individual  errors are averaged over all predictions  to obtain the overall MAE value.  More precisely, let $\mathcal{T}$ denote  the set of   ratings to be predicted, i.e., $\mathcal{T} = \{(i,j) \in [n]\times [m], R_{ij} \; \text{needs to be predicted} \}$ and let $\hat{\mathbf{R}}$ denote the  prediction matrix obtained by algorithm after factorization. Then,
\[ {\rm{MAE}} = \frac{\sum_{(i,j) \in \mathcal{T}}^{}{{|R_{ij} - \hat{R}_{ij}|}}}{|\mathcal{T}|},\]
where ${R}_{ij}$ is the real rating assigned by the user $i$ to the item $j$, and $\hat{R}_{ij}$ is the rating user $i$ would assign to the item $j$ that is  predicted by the algorithm .

The  RMSE metric is defined as: 
\[ {\rm{RMSE}} = \sqrt{\frac{\sum_{(i,j) \in \mathcal{T}}^{}{{\left(R_{ij} - \hat{R}_{ij}\right)^2}}}{|\mathcal{T}|}}.\]

The first  measure (MAE) considers every error of equal value, while the second one (RMSE)
emphasizes larger errors.   We would like to emphasize that even \textit{small improvements} in RMSE are considered valuable in the context of recommender systems. For example, the Netflix  prize competition offered a 1,000,000 reward for a reduction of the RMSE by 10\%~\cite{33333}.

\subsubsection{Metrics for evaluating the correlation of ratings with trust/distrust relations}
 As part of our experiments, we  investigate how the explicit trust/distrust relations between users in the social network are aligned with the implicit trust/distrust relations between users conveyed from the rating information.  We use  recall, Mean Average Precision (MAP)~\cite{manning2008introduction} and Normalized Discount Cumulative Gain (NDCG) to evaluate the ranking results.  \textit{Recall} is defined as the number of relevant friends divided by the total number of friends  in the social network. \textit{Precision} is defined as the number of relevant friends (trusted or distrusted) divided by the number of friends in the social network.   Given a user $u$, let $r_i$ be the relevance score of the friend ranked at position $i$, where $r_i = 1$ if the user is relevant to the $u$ and $r_i = 0$ otherwise. Then we can compute the Average Precision (AP) as
$$
{\rm{AP}} = \frac{\sum_{i}^{}{r_i \times {\rm{Precision}}@i}}{\# \text{ of relevant friends}}.
$$
MAP is the average of AP over all the users in the network. 

NDCG is a normalization of the Discounted Cumulative Gain (DCG) measure.  DCG is a weighted sum of the degree of relevancy of the ranked users. The weight is a decreasing
function of the rank (position) of the user, and therefore called
discount.  NDCG normalizes DCG by the Ideal DCG (IDCG), which is
simply the DCG measure of the best ranking result. Thus NDCG measure
is always a number in $[0,1]$. NDCG at position $k$ is defined as:
$$
{\rm{NDCG}}@k = Z_k \sum_{i=1}^{k}{\frac{2^{r_i} - 1}{\log (i+1)}}
$$
where $k$ is also called the scope, which means the number of top-ranked users presented to
the user and $Z_k$ is chosen such that the perfect ranking has a NDCG value of 1. We note that the base of the logarithm does not matter for NDCG, since constant scaling will cancel out due to normalization. We will assume it is the natural logarithm throughout this paper.

\begin{figure*}
\centering
 \includegraphics[scale=0.3]{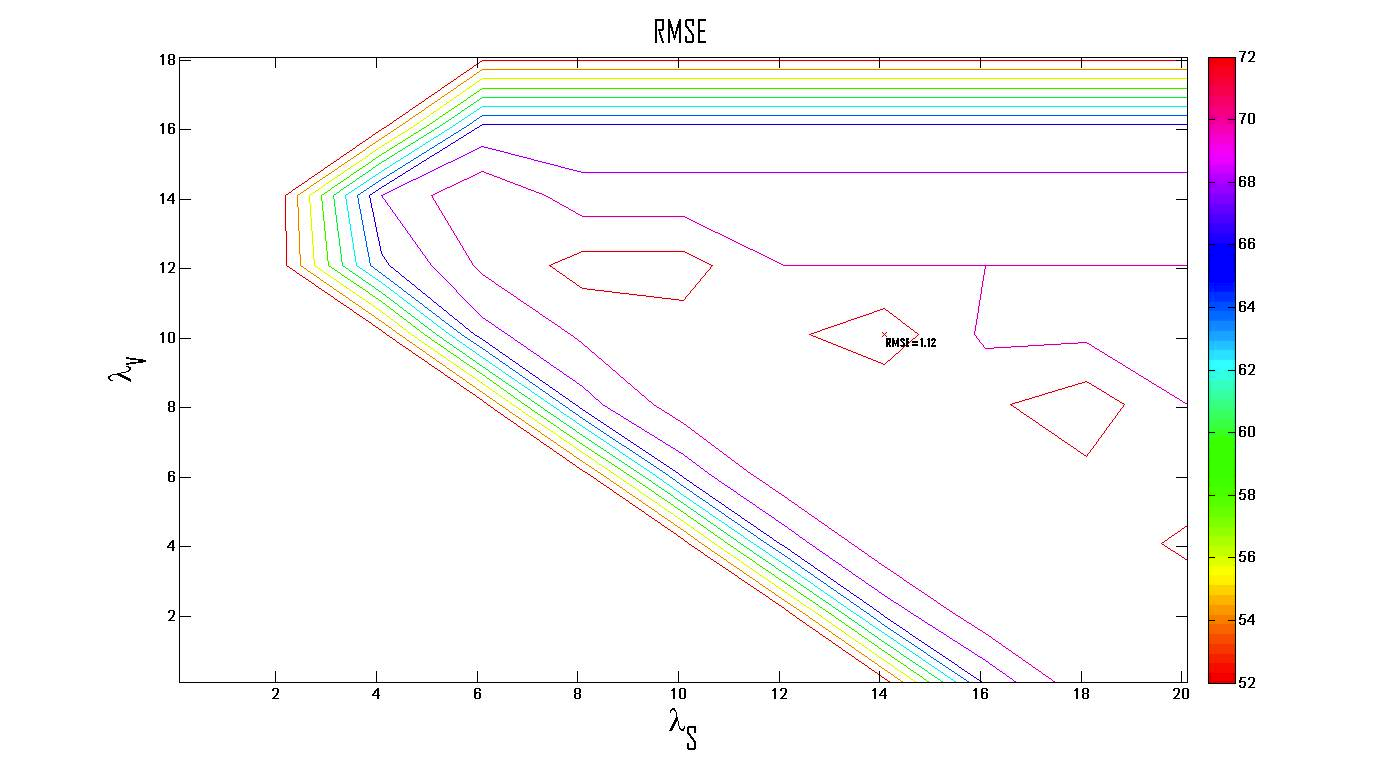}
\caption{Grid Search to find the best values for $\lambda_U$ and $\lambda_C$ on the data set with 90\% of rating information.}
\label{fig:reg}       
\end{figure*} 
\subsection{Model Selection}\label{subsub-regularization}
Tuning of parameters (a.k.a model selection in learning community) is a critical problem in most of the learning problems. In some situations, the learning performance may drastically vary with different
choices of the parameters.  There are three parameters in objective~(\ref{eqn:hing-euc}) that play very important role in the effectivity of the proposed algorithm. These are $\lambda_{U}$, $\lambda_{V}$, and $\lambda_{S}$. Between these,  $\lambda_{S}$ controls how much the proposed algorithm should incorporate the information of the social network in completing the partially observed rating matrix. In the extreme case, a very small value  for $\lambda_{S}$, the algorithm almost forgets the social information exists between the users and only utilizes the observed user-item rating matrix for factorization.  On the other hand, if we employ a very
large value for   $\lambda_{S}$, the social network information will
dominate the learning process, leading to a poorer performance.  Therefore, in order to not hurt  the recommendation performance, we need to find a reasonable value for social regularization parameter. To this end, we analyze how the combination of these parameters affect the recommendation performance.  

We conduct a grid search on the potential values of two parameters $\lambda_{S}$ and $\lambda_{V}$ to find the combination with best performance. Figure~\ref{fig:reg} shows the grid search results for these parameters on data set with 90\% of training data where the optimal prediction accuracy  is achieved at point $(14.8, 11)$ with the optimal $\mbox{RMSE} = 1.12$.  We  would like to emphasize that  we have done the cross validation for only pairs of $(\lambda_S, \lambda_V)$ and $(\lambda_S, \lambda_U)$  because,  (i) considering the grid search for the triplet $(\lambda_{S}, \lambda_{U}, \lambda_{V})$ is computationally burdensome, (ii) and our preliminary experiments showed that $\lambda_{V}$ and $\lambda_{U} $  behave similarly with respect to $\lambda_{S}$.  Based on the results reported in Figure~\ref{fig:reg}, in the remaining experiments, we set $\lambda_{S} = 14.8$,  $\lambda_{V} = 11$, and $\lambda_{U} = 13$ when the training is performed on the data set with 90\% of  rating information. We repeat the same process to find out the best  setting of regularization  parameters for other data sets with 80\%, 70\%, and 60\%  of rating data as well.

\subsection{Baseline Methods}\label{sebsec-basline}
Here we  briefly discuss the  baseline algorithms that we intend to compare the proposed algorithm. The baseline  algorithms are chosen from both types of memory-based and model-based recommender systems with different types of trust and distrust relations. In particular, we consider the following basic algorithms:

\begin{itemize}
\item \textbf{MF} (matrix factorization based recommender): this is the basic matrix factorization based recommender formulated in the optimization problem in~(\ref{eqn:mf}) which does not take the social data into account.
\item \textbf{MF+T} (matrix factorization  with trust information): to exploit the trust relations between users in matrix factorization,~\cite{44444} relied on the fact that the distance between latent features of users who trust each other must be minimized that can be formulated  as  the following objective:
$$ \min_{\Ub}\frac{1}{2} \sum_{i=1}^{n} \sum_{ j \in {\N}_{+}(i) }^{}{\D(U_{i,:}, U_{j,:})},$$
where ${\N}_{+}(i)$ is the set of users the $i$th user trusts in the social network (i.e., $S_{ij} = +1$). By employing this intuition in the basic formulation in~(\ref{eqn:mf}),~\cite{44444} solves the following optimization problem:
\begin{eqnarray*}
\min_{\Ub, \Vb} \left[\frac{1}{2}  \sum_{(i,j) \in \Omega_\Rb}^{}{\left( R_{ij} - U_{i,:}^{\top}V_{j,:}\right)^2 } + \frac{\alpha}{2}\sum_{i=1}^{n} \sum_{ j \in \N_{+}(i) }^{}{\D(U_{i,:}, U_{j,:})} + \frac{\lambda_{U}}{2} \|\Ub\|_{\rm{F}} + \frac{\lambda_{V}}{2} \|\Vb\|_{\rm{F}} \right].
\end{eqnarray*}
\item \textbf{MF+D} (matrix factorization  with distrust information): the basic intuition behind the algorithm proposed in~\cite{44444} to exploit the distrust relations is as follows: if user $u_i$ distrusts user $u_j$, then we can assume that their corresponding latent features $U_{i,:}$ and $U_{j,:}$ would have a large distance. As a result we aim to maximize the following quantity for all users:
$$ \max_{\Ub}\frac{1}{2} \sum_{i=1}^{n} \sum_{ j \in \N_{-}(i) }^{}{\D(U_{i,:}, U_{j,:})},$$
where $\N_{-}(i)$ denotes the set of users the $i$th users distrusts (i.e, $S_{ij} = -1$). Adding this term to the basic optimization problem in~(\ref{eqn:mf}) we obtain the following optimization problem:
\begin{eqnarray*}
\min_{\Ub, \Vb} \left[\frac{1}{2}  \sum_{(i,j) \in \Omega_\Rb}^{}{\left( R_{ij} - U_{i,:}^{\top}V_{j,:}\right)^2 } - \frac{\beta}{2}\sum_{i=1}^{n} \sum_{ j \in \N_{-}(i) }^{}{\D(U_{i,:}, U_{j,:})} + \frac{\lambda_{U}}{2} \|\Ub\|_{\rm{F}} + \frac{\lambda_{V}}{2} \|\Vb\|_{\rm{F}} \right].
\end{eqnarray*}
\item \textbf{MF+TD} (matrix factorization  with trust and distrust information): this algorithm stands for the algorithm proposed in the present work. We note that there is no algorithm in the literature that exploits both trust and distrust relations in factorization process simultaneously.
\item \textbf{NB} (neighborhood-based recommender): this algorithm is the basic memory-based recommender algorithm that predicts a rating of a target item $i$ for user $u$ using a combination of the ratings of neighbors of $u$ (similar users) that  already issued a rating for item $i$. Formally, 
\begin{eqnarray} \label{nb-basic}
\hat{R}_{ui} = \bar{R}_{u} + \frac{\sum_{u' \in \N(u), W_{uu'} > 0}^{}{W_{uu'} (R_{ui} - \bar{R}_{u})}}{\sum_{u' \in \N(u), W_{uu'}}^{}{W_{uu'}}},
\end{eqnarray}
where  the pairwise weight $W_{uu'}$ between pair of users $(u,u')$ is calculated by Pearson's correlation coefficient~\cite{herlocker2004evaluating}
\item \textbf{NB+T} (neighborhood  with trust information)~\cite{20004,golbeck-thesis-2005,massa2009trust}: the basic idea behind the trust based recommender systems proposed in TidalTrsut~\cite{golbeck-thesis-2005} and MoleTrsut~\cite{20004} is to limit the set of neighbors in~(\ref{nb-basic}) to the users who are trusted by user $u$. The distinguishing feature of these algorithms is the mechanism of trust propagation  to estimate the trust transitively for all the users.  By adapting~(\ref{nb-basic}) to only consider trustworthy neighbors   in predicting the new ratings we obtain:
\begin{eqnarray} \label{nb-basic-t}
\hat{R}_{ui} = \bar{R}_{u} + \frac{\sum_{u' \in \N^{*}_{+}(u), W_{uu'} > 0}^{}{W_{uu'} (R_{ui} - \bar{R}_{u})}}{\sum_{u' \in \N^{*}_{+}(u), W_{uu'} > 0}^{}{W_{uu'}}},
\end{eqnarray}
where $\N^{*}_{+}(u)$ is the set of trusted neighbors of $u$ in the social network with propagated trust relations (when there is no propagation we have $\N^{*}_{+}(u) = \N_{+}(u)$).  We note that instead of Pearson's correlation coefficient as the wighting schema, we can infer the weights  exploiting the social relation between the users. Since for the data set we consider in our experiments, the trust/distrust relations are binary values, the social based pairwise distance would be simply the hamming distance between the binary vector representation of social relations of users. For implementation details we refer to~\cite[Chapter 6]{victor2011trust}.
\item \textbf{NB+TD-F} (neighborhood  with trust information and distrust information as  filtration)~\cite{VictorCCT11,33333}: a simple strategy to use distrust relations in the recommendation is to \textit{filter} out distrusted users from the list of neighbors in predicting the ratings. As a result, we adapt~(\ref{nb-basic}) to exclude distrusted users from the users' propagated web of trust. 
\item \textbf{NB+TD-D} (neighborhood-based with trust information and integrated distrust information)~\cite{VictorCCT11,33333}: in the same spirit as the filtration strategy, we can use distrust relations to debug the trust relations.  More specifically, if user $u$ trusts user $v$,  $v$ trusts $w$,   and $u$ distrusts  $w$, then the latter distrust relation contradicts the propagation of the trust from $u$ to $w$ and can be excluded from the prediction. In this method distrust is used to \textit{debug} the trust relations. 
\end{itemize}
\begin{table*}[t]
\centering
\caption{\ct{The consistency of implicit and explicit trust relations in the data set for different ranges of ratings measured in terms of NDCG, recall, and MAP.}\label{table:com-social-trsut}}{%
\begin{tabular}{|l||c|c||c|c|c||c|}
\hline
\# of Ratings & NDCG@10 & NDCG@20 & Recall@10 & Recall@20 & Recall@40 & MAP \\
\hline \hline 
0-20 & 0.083 & 0.078 & 0.054 & 0.092 & 0.156 & 0.140  \\ \hline
21-40 & 0.108 & 0.103 & 0.080 & 0.125 & 0.198 & 0.190  \\ \hline 
41-60 & 0.117 & 0.112 & 0.083 & 0.128 & 0.225 & 0.208  \\ \hline
61-80 & 0.120 & 0.117 & 0.088 & 0.132 & 0.230 & 0.230   \\ \hline 
$ \ge 81\;$& 0.135 & 0.126 & 0.091 & 0.151  & 0.253  & 0.244  \\ \hline
\end{tabular}}
\end{table*}
\begin{table*}[t]
\centering
\caption{\ct{The consistency of implicit and explicit distrust relations in the data set for different ranges of ratings measured in terms of NDCG, recall, and MAP.}\label{table:com-social-distrsut}}{%
\begin{tabular}{|l||c|c||c|c|c||c|}
\hline
\# of Ratings & NDCG@10 & NDCG@20 & Recall@10 & Recall@20 & Recall@40 & MAP \\
\hline \hline 

0-20 & 0.065 & 0.057 & 0.045 & 0.071 & 0.132 & 0.130  \\ \hline 
21-40 & 0.071 & 0.068 & 0.060 & 0.077 & 0.140 & 0.134  \\ \hline 
41-60 & 0.082 & 0.072 & 0.075 & 0.085 & 0.158 & 0.152  \\ \hline 

61-80 & 0.089  & 0.078 & 0.081 & 0.105 & 0.164 & 0.160  \\ \hline 
$ \ge 81\;$ & 0.104 & 0.096 & 0.087 & 0.125 & 0.191 & 0.183   \\ \hline

\end{tabular}}
\end{table*}
\subsection{On the Consistency of  Social Relations and Rating Information}\label{sec:emp-cons}
As already mentioned, the Epinions website allows users to write reviews about
products and services and  to rate reviews written by other users. Epinions also allows users to define their web of trust, i.e. "reviewers whose reviews and ratings have been consistently
found to be valuable" and their block list, i.e. "reviewers whose reviews are found to be consistently
 inaccurate or not valuableÓ.  Different intuitions on interpreting these
social information will result in different  models. The  main rational behind incorporating  trust and distrust relations in recommendation process is  to take the trust/distrust relations between users in the social network as the level of agreement between ratings  assigned to reviews by users~\footnote{In the literature the similarity between users conveyed from the  rating information issued by users and the direct relation in the social network  are  usually referred to as the \textit{implicit} and the \textit{explicit} trust, respectively.}.  Therefore, investigating the consistency or alignment between user ratings (implicit trust) and trust/distrust relations in the social network (explicit trsut) become an important issue. 

Here, we aim to empirically investigate whether or not there is a correlation between  a user's current trustees/friends  or distrusted friends and the ratings that user would assign to reviews issued by his neighbors. Obviously,  if there is no correlation between social context of a user and his/her ratings to reviews written by his neighbors, then the social structure does not provide any advantage to the rating information. On the other hand, if there exists such a correlation,  then the social context could be supplementary information to compensate for the lack of rating information   to boost the accuracy of recommendations.

The consistency of trust relations and rating information issued by users on the reviews written by his trustees has been analyzed  in~\cite{ziegler2007investigating,guo2014ratings}. However,~\cite{ziegler2007investigating} also claimed that social trust (i.e., explicit trust) and similarity between users based on their issued ratings (i.e., implicit trust) are not the same, and can be used complementary. According to \cite{ma2013experimental}, when comparing implicit social information with explicit social information, the performance of using implicit information is slightly worse.  We further investigate the same question about the consistency of distrust relations and ratings issued by users to their distrusted neighbors. The positive answer to this question can be interpreted as follows. Given that user $u$ is interested in item $i$, the chances that $v$, trusted (distrusted) by $u$, also likes this item $i$ is much higher (lower) than for user $w$ not explicitly trusted (distrusted) by $u$.

To measure the similarity between users, there are several methods we can borrow in the literature. In this paper, we adopt the most popular approach that is referred to as  Pearson correlation coefficient (PCC) $\mathcal{P}: \U \times \U \mapsto [-1,+1]$~\cite{breese1998empirical,massa2009trust}, which is defined as:
$$ \mathcal{P}(u, v) = \frac{\sum_{i=1}^{m}{(R_{ui}-\bar{R}_u)(R_{vi}-\bar{R}_v)}}{\sqrt{\sum_{j=1}^{m}{(R_{ui} - \bar{R}_u)^2} \times \sum_{j=1}^{m}  {(R_{vi} - \bar{R}_v)^2}} }, \forall u, v \in \U,$$
where $\bar{R}_u$ and $\bar{R}_v$ are  the average of ratings issued by users $u$ and $v$, respectively.  The PCC measures the extent to which there is a linear relationship between the rating behaviors of the two users, the extreme values being  -1 and 1. The similarity of two users becomes negative when users have completely diverging ratings.  We note that this quantity can be considered as the implicit trust between users that is conveyed via ratings given by users. 
 
\begin{table*}[t]
\centering
\caption{\ct{The alignment rate of users in  establishing trust/distrust relationships with future users in the social network based on the majority vote of their current trusted/distrusted friends. The number of trusted friends (+) and distrusted friends (-) are denoted by $n_{+}$ and $n_{-}$, respectively. Here $u$ denotes the current user and $w$ stands for a future user in the network.}\label{table:com-social-cons}}{%
\begin{tabular}{|l|c|c|c|}
\hline
Setting & Type of Relation ($u \leadsto  w$) & \% of Relations & Alignment Rate (\%) \\
\hline \hline 
$n_{+} > n_{-}$ & + & 48.80 & 92.09 \\
& - & 2.54 & 8.15 \\ \hline
$n_{+} < n_{-}$ & + & 1.15 &  17.88\\
& - & 8.02 & 83.42  \\ \hline
$n_{+}  = n_{-} > 0$ or $n_{+}  = n_{-}  =  0$ & + & 39.49 & -  \\  \hline \hline
\end{tabular}}
\end{table*}

To conduct this set of experiments, we first group all the users in the training data set based on the number of  ratings, and then measure the prediction accuracies of different user groups. Users are grouped into five classes: "[1, 20)", "[20, 40)", "[40, 60)", "[60, 80)",  and "$>81$ ".  In order to have a comprehensive view of the ranking performance, we present the NDCG, recall and MAP scores of trust and distrust alignments on the Epinions data set in Table~\ref{table:com-social-trsut} and Table~\ref{table:com-social-distrsut}, respectively.  We note that the data set we use in our experiments only contains bivalent trust values, i.e., -1 and +1, and it is not possible to have an ordering on the list of friends (timestamp of relations would be an option to order the friends but unfortunately it is not available in our data set). To compute the NDCG, we use the ordering of trusted/distrusted friends which yields the best value.

On the positive side,  we observe a clear trend of alignment between ratings  assigned by a user and the type of relation he has made in the social network.  This observation coincides with our intuition. Overall, when more ratings are observed for a user, the similarity calculation process will find more accurate similar or dissimilar neighbors for this user since we have more information to represent or interpret this user. Hence, by increasing the number of ratings, It is
conceivable  from the results in Tables~\ref{table:com-social-trsut} and~\ref{table:com-social-distrsut} that the alignment between implicit and explicit neighbors becomes better. By comparing the results in Tables~\ref{table:com-social-trsut} and~\ref{table:com-social-distrsut} we can see that trust relations are slightly better aligned than the distrust relations. 

On the negative side, the results show that the NDCG on both types of relations is small.  One explanation for this phenomenon is that the Epinions data set is not tightly bound to a specific application. For example, a user may trust or distrust anther user based on his/her comments on a specific product but they might have similar taste on other products. Furthermore,  compared to other data sets such as FilmTrusts, the Epinions data set is very sparse data set, and consequently it is relatively inaccurate to rely on the rating information to compute the implicit trust relations. Finally,  our approach to distinguish  trust/distrust lists from the rating information is limited by the  PCC trust metric we have utilized. We conjecture that better trust  metrics that is able to exploit other side information such as time and interactional information would be helpful in distinguishing implicit trusted/distrusted friends, leading to better alignment between implicit and explicit trust relations.

We also conduct experiments to evaluate the consistency of social network only based on the trust/distrust relations between users.  In particular,  we investigate to what extent a users' relations are aligned with the opinion of his/her  neighbors in the social network. More specifically, let $u$ be a user who is about to make a trust or distrust relation to another user $v$. We assume that $n_{+}$ number of $u$'s neighbors trust $v$ and $n_{-}$ number of $u$'s neighbors distrust $v$. We note that in the real data set the distrust relations are hidden.  To conduct this set of experiments, we randomly sample 30\% of the relations from the social network and use the remaining 70\% to predict the type of sampled relations~\footnote{A more realistic way would be to use the timestamp of relations to create the training and test sets.} by \textit{majority voting}.  

Table~\ref{table:com-social-cons} shows the results on the consistency of social relations. We observe that in all cases there is an alignment between the opinions of users' friends and his/her own relation (92.09\% and 83.42\% when the  majority of friends trust and distrust the target user, respectively).   This might be due to social influence of people on social network, however, it is hard to justify the existence of such a correlation in Epinions data set which includes reviews for diverse set of products and taste of users.  One interesting observation from the results reported in Table~\ref{table:com-social-cons} is the case where the number of distrusted users dominates the number of trusted users (i.e., $n_{-} > n_{+}$). While the distrust relations are private to other users, but we can see that there is a significant alignment between users's relation type and his distrusted friends. 

\begin{table*}[t]
\centering
\caption{\ct{The accuracy of prediction of matrix factorization with three different methods measured in terms of MAE and RMSE errors. The parameter $k$ represents the number of latent features in factorization}.\label{table:res}}{%
\begin{tabular}{|c||c|l|c|c|c| c|}
\hline
$k$ & $\%$  of Training & Measure & \textbf{MF} & \textbf{MF+T} & \ct{\textbf{MF+D}} & \textbf{MF+TD} \\
\hline \hline 

10 & 60\% & MAE & 0.9813 $\pm$ 0.042 & 0.8561 $\pm$ 0.032 & 0.9720 $\pm$ 0.038 & 0.8310 $\pm$ 0.016 \\
 &  & RMSE & 1.6050 $\pm$ 0.032 & 1.4125 $\pm$ 0.022 & 1.5036 $\pm$ 0.040 &1.2294 $\pm$ 0.086 \\
 \hline 

 & 70\% & MAE & 0.9462 $\pm$ 0.083  & 0.8332 $\pm$ 0.092 &0.9241 $\pm$ 0.012 & 0.8206 $\pm$ 0.023 \\
 &  & RMSE & 1.5327 $\pm$ 0.032 & 1.2407 $\pm$ 0.063 & 1.4405 $\pm$ 0.023 & 1.1562 $\pm$ 0.043 \\
 \hline 

  & 80\% & MAE & 0.9150$\pm$ 0.022  & 0.8206 $\pm$ 0.041 & 0.8722 $\pm$ 0.034 &  0.8113 $\pm$ 0.032 \\
 &  & RMSE & 1.3824 $\pm$ 0.032 & 1.1906 $\pm$ 0.042 & 1.3155 $\pm$ 0.026 & 1.1061 $\pm$ 0.021 \\
 \hline 

  & 90\% & MAE & 0.8921 $\pm$ 0.025 & 0.8158 $\pm$ 0.016  & 0.8736 $\pm$ 0.053 & 0.8025 $\pm$ 0.014 \\
 &  & RMSE & 1.2166 $\pm$ 0.017 & 1.1403 $\pm$ 0.027 & 1.1869 $\pm$ 0.049 & 1.0872 $\pm$ 0.020 \\
 \hline

\hline \hline
20 & 60\% & MAE & 0.9972 $\pm$ 0.016 & 0.8431 $\pm$ 0.018 &  0.9746 $\pm$ 0.060 & 0.8475 $\pm$ 0.012 \\
 &  & RMSE & 1.6248 $\pm$ 0.014 & 1.3904 $\pm$ 0.042 & 1.5423 $\pm$ 0.046 & 1.1837 $\pm$ 0.023 \\
 \hline 

  & 70\% & MAE & 0.9688 $\pm$ 0.019 & 0.8342 $\pm$ 0.062 & 0.9350 $\pm$ 0.022 & 0.8290  $\pm$ 0.034 \\
 &  & RMSE & 1.5162 $\pm$ 0.016 & 1.2722 $\pm$ 0.027 & 1.4540 $\pm$ 0.075 & 1.1452 $\pm$ 0.016 \\
 \hline 

  & 80\% & MAE & 0.9365 $\pm$ 0.025 & 0.8172 $\pm$ 0.011 & 0.8705 $\pm$ 0.016 & 0.8129 $\pm$ 0.025 \\
 &  & RMSE & 1.4081 $\pm$ 0.015 & 1.1853 $\pm$ 0.023 & 1.3591 $\pm$ 0.073 & 1.1049 $\pm$ 0.082 \\
 \hline 

  & 90\% & MAE & 0.9224 $\pm$ 0.016 & 0.8128  $\pm$ 0.021 & 0.8805 $\pm$ 0.032  & 0.8096 $\pm$ 0.010 \\
 &  & RMSE & 1.2207 $\pm$ 0.0 18& 1.1402 $\pm$ 0.026 & 1.1933 $\pm$ 0.028& 1.0851 $\pm$ 0.011 \\
 \hline
 \hline
\end{tabular}}
\end{table*}
\subsection{On the Power of  Utilizing Social Relationships}
We now turn to investigate the effect of utilizing  social relationships between users on the  accuracy of recommendations  in \textit{factorization-based} methods. In other words, we would like to   experimentally evaluate whether incorporating distrust can indeed enhance the trust-based recommendation process.   To this end, we run four different  \textbf{MF} (i.e., pure matrix factorization based algorithm), \textbf{MF+T} (i.e.,  matrix factorization with only trust relationships), \textbf{MF+D} (i.e.,  matrix factorization with only distrust relationships), and \textbf{MF+TD} (i.e., the algorithm proposed here) algorithms   on the data set.  We run the algorithms with  $k = 10$ and $k = 20$ latent vector dimensions.  As mentioned earlier, different amount of training data 90\%, 80\% , 70\% and 60\%   has been used to create four different training sets that are increasingly sparse but the social network remains the same in all of them. We evaluate all algorithms by both MAE and RMSE measures.

Table~\ref{table:res} shows the MAE and RMSE errors for the four sampled data sets.  First, as we expected, the performance of all learning algorithms improves with an increasing number of training data. It is also not surprising to see that the   \textbf{MF+T}, \textbf{MF+D}, and \textbf{MF+TD} algorithms which exploit social side information  perform better than the pure matrix factorization based  \textbf{MF} algorithm. Second, the proposed algorithm outperforms all   other baseline  algorithms  for all the cases, indicating that it is effective to incorporate    both types of social side information in recommendation. This result by itself indicates that besides trust relationships in the social network, the distrust information is also a rich source of information and can be utilized in recommendation algorithms. We note that distrust information needs to be incorporated carefully as its nature is totally different from trust information. Finally, it is noticeable that the \textbf{MF+T} outperforms the \textbf{MF+D} algorithm due to huge number of trust relations to distrust relations in our data set. It is also remarkable that users  are more likely to be influenced by their friends to make trust relations than the distrust relations due to the private nature of distrust relations in Epinions data set. This might lead us to believe that  distrust relations have better quality than trust relations which  requires a deeper investigation to be verified.  
\subsection{\ct{Comparison to Baseline Algorithms}}
Another  question that is worthy of investigation  is how state-of-the-art approaches  perform compared to the method proposed in this paper. To this end, we compare the performance of the \textbf{MF-TD} algorithm with the  baseline algorithms introduced in Subsection~\ref{sebsec-basline}. Table~\ref{table:com-other} contains the results of our experiments with eight different algorithms on the data set with 90\% of rating data.  The second column in the table represents the configuration of parameters used by each algorithm. 

 When we utilize trust/distrust relations in neighborhood-based   algorithms, a crucial decision we need to make is  to which level the propagation must be performed (no propagation corresponds to the single level  propagation which only includes direct neighbors).  Let $p$ and $q$ denote the level of propagation for trust and distrust relations, respectively. Let us first consider the \textit{trust} propagation to decide the value of $p$. We  note that there is a tradeoff between accuracy and the level of trust propagation: the longer  propagation levels results in  less accurate trust predictions. This is due the fact that when we use longer propagation levels,  the further away we are heading from each user, and consequently  decrease the confidence on the predictions.   Obviously this  affects the accuracy of the recommendations significantly. As a result, for the trust propagation we only consider single level propagation by choosing $p=1$ (i.e, $\N^*_{+} = \N_{+}$). We also note that since in the Epinions data set  a user can not simultaneously trust and distrust another user, in the neighborhood-based method with distrust relations,  the debugging only makes sense for propagated information. Therefore, we perform a three level distrust propagation ($q = 3$) to constitute the set of distrusted users for each users.  We note that the longer the propagation levels, the more often distrust evidence can be found for a particular user, and hence the less  neighbors will be left to participate in the recommendation process. For factorization based methods, the value of regularization parameters, i.e., $\lambda_U$, $\lambda_V$, and $\lambda_S$, are determined by the procedure discussed in Subsection~\ref{subsub-regularization}.

The results of Table~\ref{table:com-other}  reveal some interesting conclusions as summarized below:
\begin{itemize}
\item From Table~\ref{table:com-other}, we can observe that  for factorization-based methods, incorporating  trust or distrust information boost the performance of recommendation in terms of both accuracy measures. This demonstrates the advantages of trust and distrust-aware recommendation algorithms. We also can see  that both \textbf{MF+T} and \textbf{MF+D} perform better than the non-social \textbf{MF} but the performance of \textbf{MF+T} is significantly better than \textbf{MF+D}. As discussed before, this observation does not indicate that the trust relations are more beneficial than the distrust relations as in our data set only $16.7\%$ of relations are distrust relations. The \textbf{MF+TD} algorithm that is able to employ both types of relations is significantly better than other algorithms that demonstrates the advantages of proposed method to utilize trust and distrust relations.

\item Looking at the results reported in Table~\ref{table:com-other}, it can immediately be noticed that the incorporation of trust and distrust information in neighborhood-based methods decreases the prediction error but the improvement is not as significant as the factorization based methods. We note that for the \textbf{NB+T} method with longer levels of propagation ($p = 2, 3$),  our experiments revealed that the accuracy remains almost same or gotten worse  on both  MAE and RMSE measures and this is why we only report the results only for $p=1$. In contrast, for distrust propagation we found out that $q=3$ has a visible impact on the performance of both filtering and debugging methods. We would  like to emphasize that for longer levels of distrust propagation in Epinions data set, i.e., $q > 4$, we found that  the size of the set of distrusted users $\N^*_{-}(\cdot)$  becomes large for most of users which degrades the prediction accuracy. We also observe another interesting result about the performance of \textbf{NB+TD} method with filtering and debugging strategies. We found that although filtering generates slightly better predictions,  \textbf{NB+TD-F} performs almost as good as the  \textbf{NB+TD-D}  method. Although this observation does not suggest any of these methods as the method of choice in incorporating distrust, we believe that the accuracy might differ from data set to data set and it strongly depends on the propagation/aggregation strategy.  
\item Considering the results for both model-based and memory-based methods in Table~\ref{table:com-other}, we  can conclude few interesting observations. First, we notice that factorization-based methods with trust/distrust  information perform better than the neighborhood based methods. Second, the incorporation of trust and distrust relations in matrix factorization has significant improvement compared to improvement  achieved by  memory-based methods.  Although the type of filtration or debugging strategy  could significantly affect the accuracy of incorporating distrust in memory-based methods, but the main shortcoming of these methods comes from the fact that these algorithms   somehow exclude the influence of distrusted users from the rating prediction. This stands in stark contrast to the model proposed in this paper that   ranks the neighbors based on the type of relation.  This observation necessitates to devise better algorithms for propagation and aggregation of trust/distrust information in memory-based methods. 
\end{itemize}
\begin{table*}[t]
\centering
\caption{\ct{Comparison with other popular methods. The reported values are the MAE and RMSE on the data set with 90\% of rating information. The values of parameters for each specific algorithm is included in the second column.}\label{table:com-other}}{%
\begin{tabular}{|l||l|c|c|}
\hline
Method & Parameter (s) & MAE & RMSE \\
\hline \hline 
\textbf{MF} & $k = 10$ and $\lambda_U = \lambda_V = 5$ & 0.8921 & 1.2166 \\ \hline
\textbf{MF+T} & $k = 10$, $\lambda_U = \lambda_V = 5$ , and $\alpha = 1$& 0.8158 & 1.1403 \\ \hline 
\textbf{MF+D} & $k = 10$, $\lambda_U = \lambda_V = 5$ , and $\beta = 10$ & 0.8736 & 1.1852\\ \hline
\textbf{MF+TD}  & $k = 10$, $\lambda_U = 13$, $\lambda_V = 11$ , and $\lambda_S = 14.8$ & 0.8025 & 1.0872\\ \hline \hline 
\textbf{NB} & & 0.9381 & 1.5275 \\ \hline
\textbf{NB+T} & $p = 1$   &0.8904 & 1.3455\\ \hline 
\textbf{NB+TD-F} & $p=1$ and $q=3$&0.8692 & 1.2455\\ \hline
\textbf{NB+TD-D} & $p=1$ and $q=3$ & 0.8728 & 1.2604\\ \hline  \hline
\end{tabular}}
\end{table*}
\begin{table*}[t]
\centering
\caption{The accuracy of handling cold-start users and the effect of social relations. The number of leant features in this experiments is set to  $k=10$. The first column shows the number of cold-start users sampled randomly from all users in the data set. For the cold-starts users all the ratings have been  excluded from the training data and used in the evaluation of three different algorithms.\label{table:cold}}{%
\begin{tabular}{|c|l|c|c|c|c|}
\hline
 $\%$ of Cold-start  Users & Measure & \textbf{MF} & \textbf{MF+T} & \ct{\textbf{MF+D}} &  \textbf{MF+TD}\\
\hline \hline 

30\% & MAE & 0.9923   & 0.8824  & 0.9721& 0.8533  \\
  & RMSE & 1.7211   & 1.5562  & 1.6433 &1.4802  \\
 \hline 

20\% & MAE & 0.9812   & 0.8805 & 0.9505 &0.8472  \\
  & RMSE & 1.7088  & 1.4339  & 1.6250 &1.2630  \\
 \hline 

 10\% & MAE & 0.9334  & 0.8477  & 0.9182 & 0.8322  \\
  & RMSE & 1.4222  & 1.3782  & 1.4006 & 1.2655  \\
 \hline 

  5\% & MAE & 0.9134  & 0.8292  &  0.8633 & 0.8280  \\
   & RMSE & 1.3852 & 1.2921  & 1.3255 & 1.2888  \\
 \hline \hline
\end{tabular}}
\end{table*}
\subsection{Handling Cold-start Users by Social Side Information}

In this subsection, we demonstrate the use of social network   to further illustrate the  potential of  proposed framework and the relevance of incorporating side information. To do so, as another  set of our experiments, we  intend to examine the performance of proposed algorithm on clod-start users. Addressing cold-start users (i.e., users with few ratings or new users) is a very important for the success of  recommender systems due to huge number of this type of  users in many real world systems.  As a result, handling cold-start users is one the main challenges in existing systems. To evaluate different algorithms we randomly select $30\%$, $20\%$, $10\%$, and $5\%$ as the cold-start users. For cold-start users, we do not include any rating in  the training data and consider all the ratings made by cold-start users as testing data.

Table~\ref{table:cold} shows the performance of above mentioned algorithms.  As it is clear from the Table~\ref{table:cold}, when the number of cold-start users is low with respect to the total number of users, say $5\%$ of total users, the affect  of distrust relationships is negligible in prediction accuracy. But, when the number of cold-start users is high, exploiting the trust and distrust relationships significantly improve the performance of recommendation. This result is interesting as it reveals that the lack of rating information for cold-start and new users can be alleviated by incorporating the social relations of users, and in particular both trust and distrust relationships. 
\begin{table*}[t]
\centering
\caption{\ct{The accuracy of proposed algorithm  on a data set with 390257 ($\approx 90\%$) trust relations sampled uniformly at random from all trust relations with varied number of distrust relations.  The learning is performed based on  $90\%$ of all ratings with $k=10$ as the dimension of latent features}.\label{table:com-base}}{
\begin{tabular}{|l|l|c|l|c|}
\hline
Method & $\#$ of Trust Relations & $\#$ of Distrust Relations & Measure & Accuracy \\
 \hline \hline

\textbf{MF+TD}   & 433,619 ($\approx 90\%$) &   9,682 ($\approx 10\%$)  & MAE &   0.8803 $\pm$  0.051      \\
  & &  & RMSE &  1.2166 $\pm$ 0.028    \\
 \hline 

 & &    19,364 ($\approx 20\%$)  & MAE &   0.8755 $\pm$ 0.033        \\
  & &  & RMSE &  1.1944 $\pm$ 0.042    \\
 \hline

 & &    29,047 ($\approx 30\%$)  & MAE &  0.8604  $\pm$   0.036     \\
  & &  & RMSE &  1.1822 $\pm$  0.081   \\
 \hline 

 & &   38,729 ($\approx 40\%$)  & MAE &  0.8431 $\pm$  0.047      \\
  & &  & RMSE &  1.1706$\pm$    0.055 \\
 \hline 

  & & 48,411 ($\approx 50\%$)  & MAE &  \textbf{0.8165}$\pm$   0.056     \\
  & & &  RMSE &  \textbf{1.1425}$\pm$  0.091   \\
 \hline 
  & & 58,093 ($\approx 60\%$)  & MAE &  0.8130$\pm$  0.035      \\
  & & &  RMSE &  1.1380$\pm$  0.046   \\
 \hline 
  & & 67,776 ($\approx 70\%$)  & MAE &  0.8122 $\pm$    0.041    \\
  & & &  RMSE & 1.1306  $\pm$ 0.042    \\
 \hline 

  & & 77,458 ($\approx 80\%$)  & MAE &  0.8095 $\pm$  0.036      \\
  & & &  RMSE &  1.1290 $\pm$ 0.085    \\
 \hline 

  & & 87,140 ($\approx 90\%$)  & MAE & 0.8061  $\pm$ 0.044       \\
  & & &  RMSE &  1.1176 $\pm$  0.067   \\
 \hline

  & & 96,823 ($= 100\%$)  & MAE &  0.8050 $\pm$  0.052       \\
  & & &  RMSE &  1.1092 $\pm$ 0.063    \\
\hline \hline

\textbf{MF+T} & 481,799 ($= 100\%$) & 0 & MAE & 0.8158 $\pm$ 0.016\\
 & &  & RMSE & 1.1403 $\pm$ 0.027\\
\hline \hline
\end{tabular}}
\end{table*}
\subsection{Trading Trust for Distrust Relationships}

We also compare the   potential  benefit of trust relations to distrust relations in the proposed algorithm. More specifically, we would like to see in what extent the distrust relations can compensate for the lack of trust relations. We run the proposed algorithm with the subset of trust and distrust relations and compare it to the algorithm which only utilizes all of the trust relations. To setup this set of experiments, we randomly sample a subset of trust relations and  gradually increase the amount of distrust relations to see when the effect of distrust information compensate the effect of missed trust relations.

We sample  433,619 (approximately $90\%$)  trust relations from the total 481,799  trust relations and vary the number of distrust relations and feed to the proposed algorithm.  Table~\ref{table:com-base} reports the accuracy of proposed algorithm  for different number of  distrust relations  in the data sets. All these samplings have been done uniformly at random. We use $90\%$ of all ratings for training and the remaining $10\%$ for evaluation, and set the dimension of latent features to $k=10$.  As it can be concluded from Table~\ref{table:com-base}, when we feed the proposed algorithm \textbf{MF+TD} with $90\%$ of trust and $50\%$ of the  distrust relations, it reveals very similar behavior to the trust-enhanced matrix factorization based method \textbf{MF+T}, which only utilizes all the trust relations in factorization.  This result is interesting in the sense that the distrust information between users is as important as the trust information (we note that in this  scenario the number trust relations excluded from the training is  almost same as the number of distrust relations included).  By increasing the number of distrust relations we can observe that the accuracy of recommendations increases as expected. In summary, this set of experiments validates that incorporating distrust relations can indeed enhance the trust-based recommendation process and could be considered as a rich source of information  to be exploited.

\begin{figure}[t]
\centering
\subfigure[$60 \%$ of Training Data]{\includegraphics[scale=0.35]{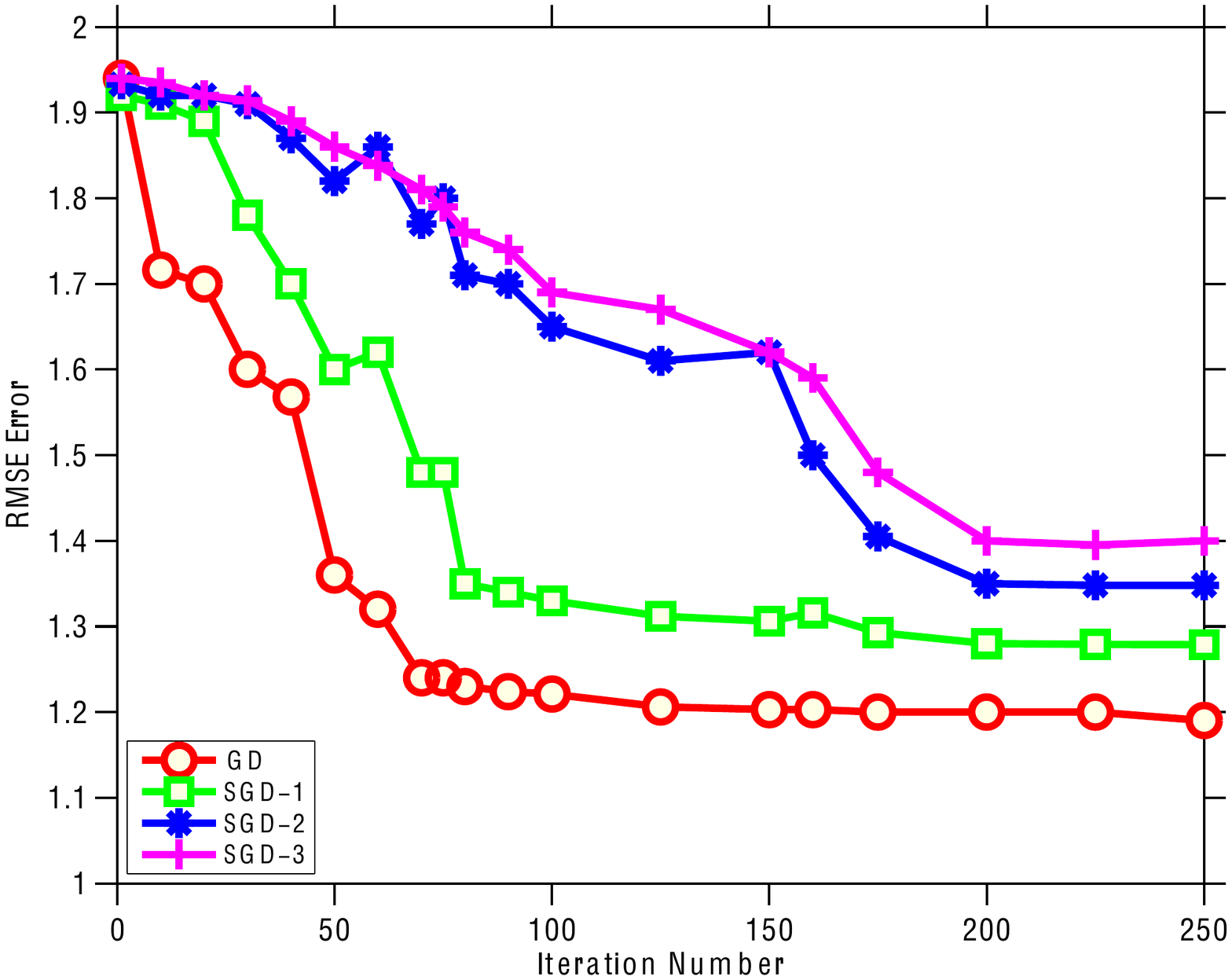}}\hspace*{-0.1in}
\subfigure[$70 \%$ of Training Data]{\includegraphics[scale=0.35]{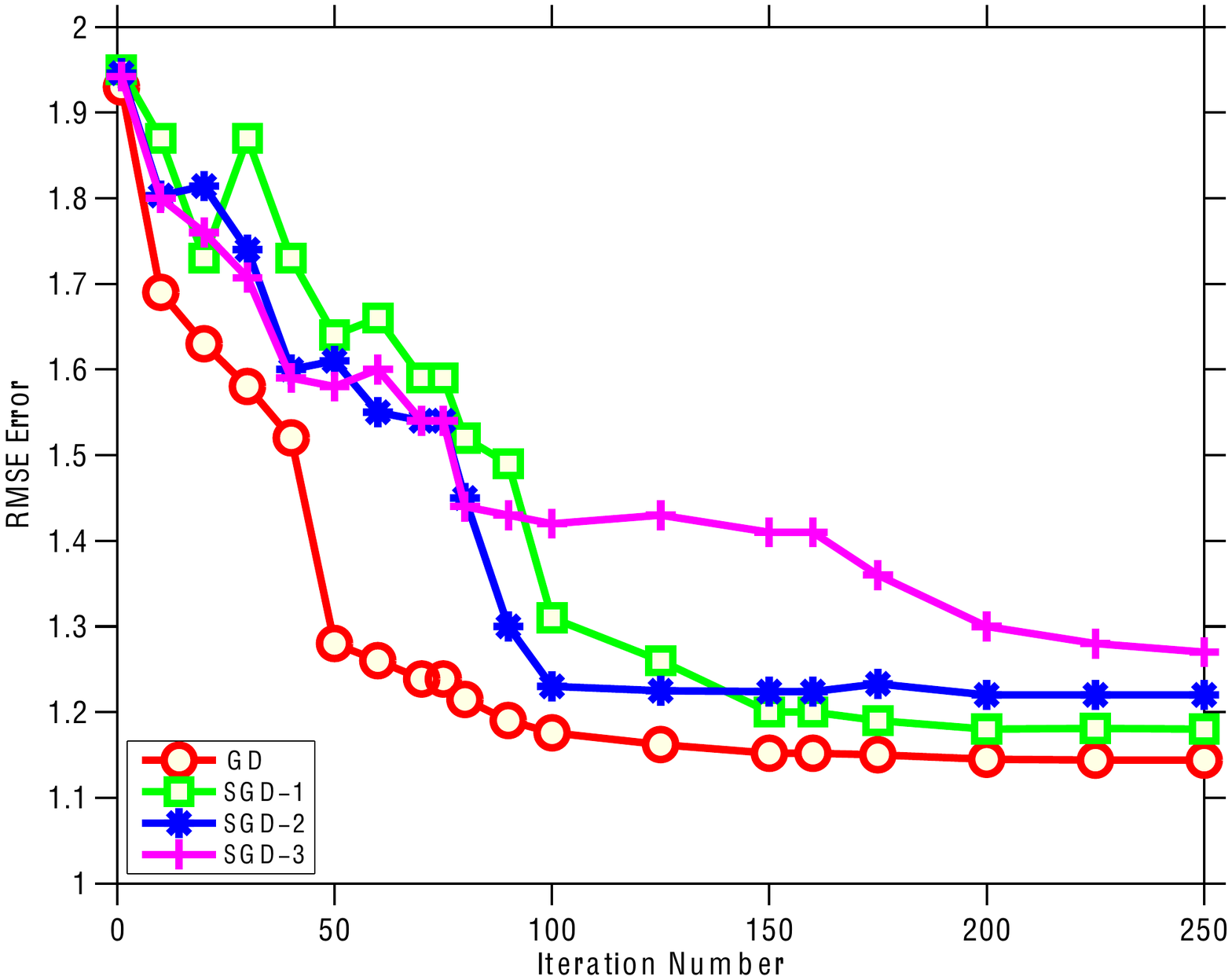}}
\centering
\subfigure[$80 \%$ of Training Data]{\includegraphics[scale=0.35]{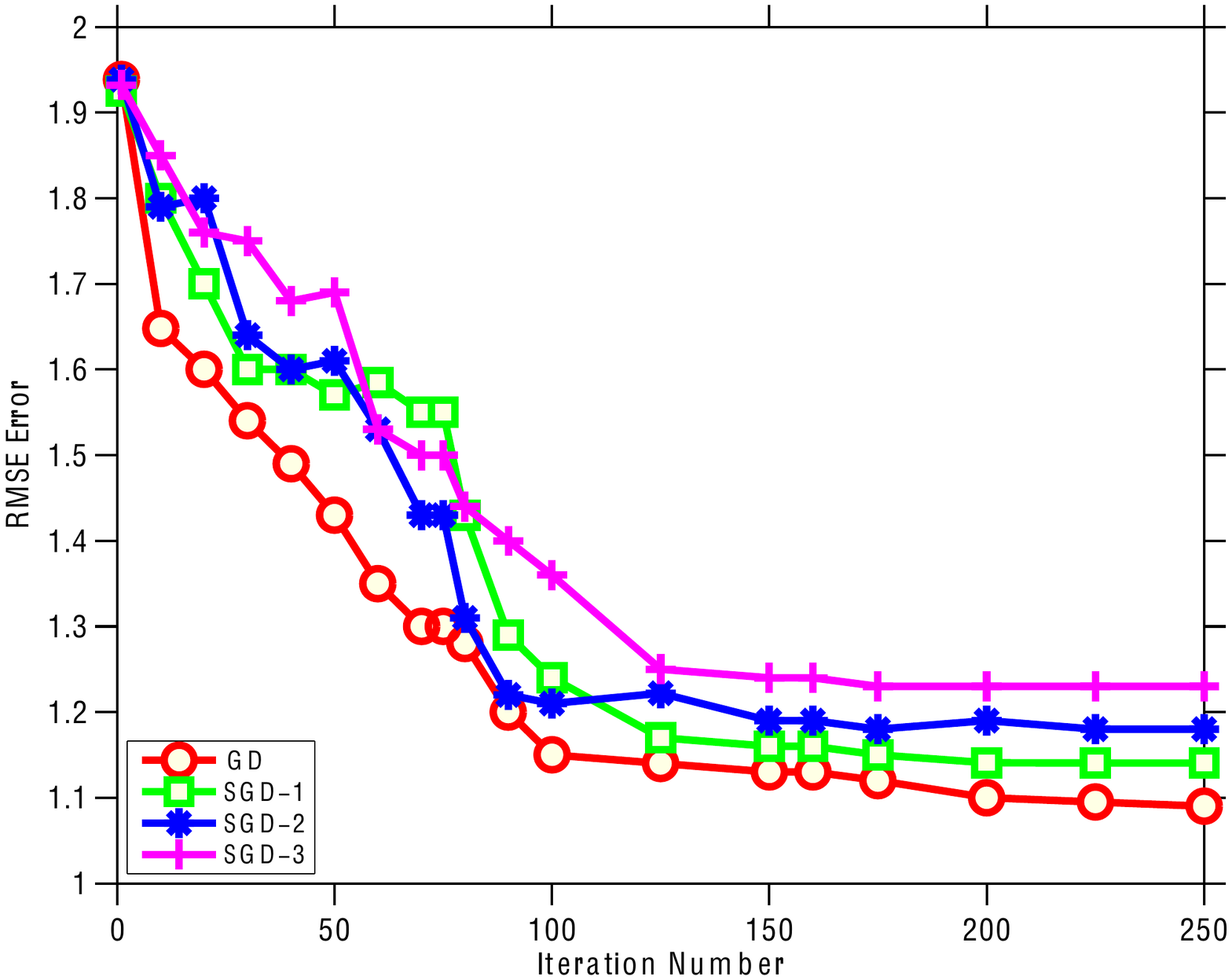}}\hspace*{-0.1in}
\subfigure[$90 \%$ of Training Data]{\includegraphics[scale=0.35]{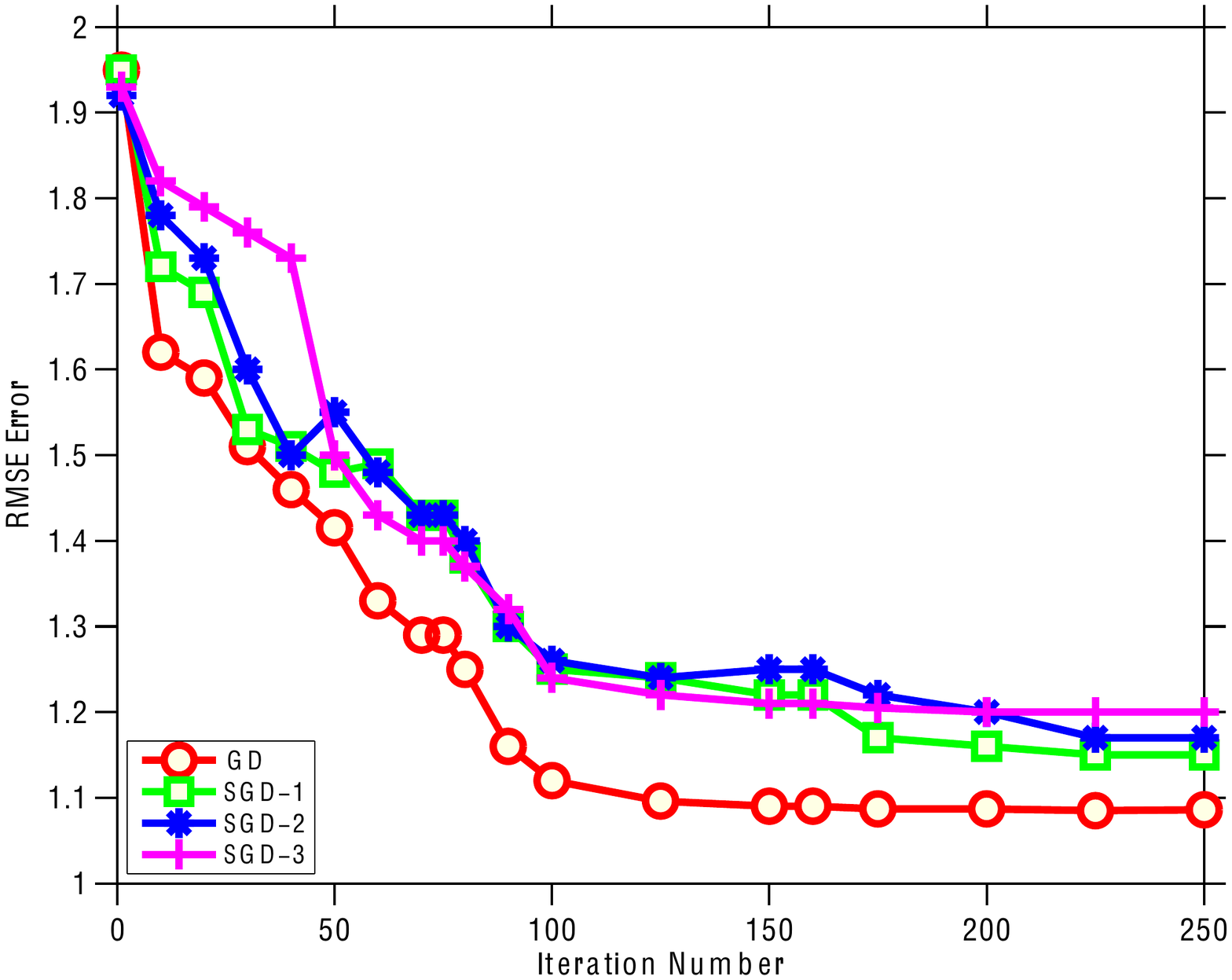}}
\caption{Comparison of accuracy of prediction in terms of RMSE with GD and SGD with three varied batch sizes.}
\label{fig:rmse}
\end{figure}
\begin{figure}[t]
\centering
\subfigure[$60 \%$ of Training Data]{\includegraphics[scale=0.35]{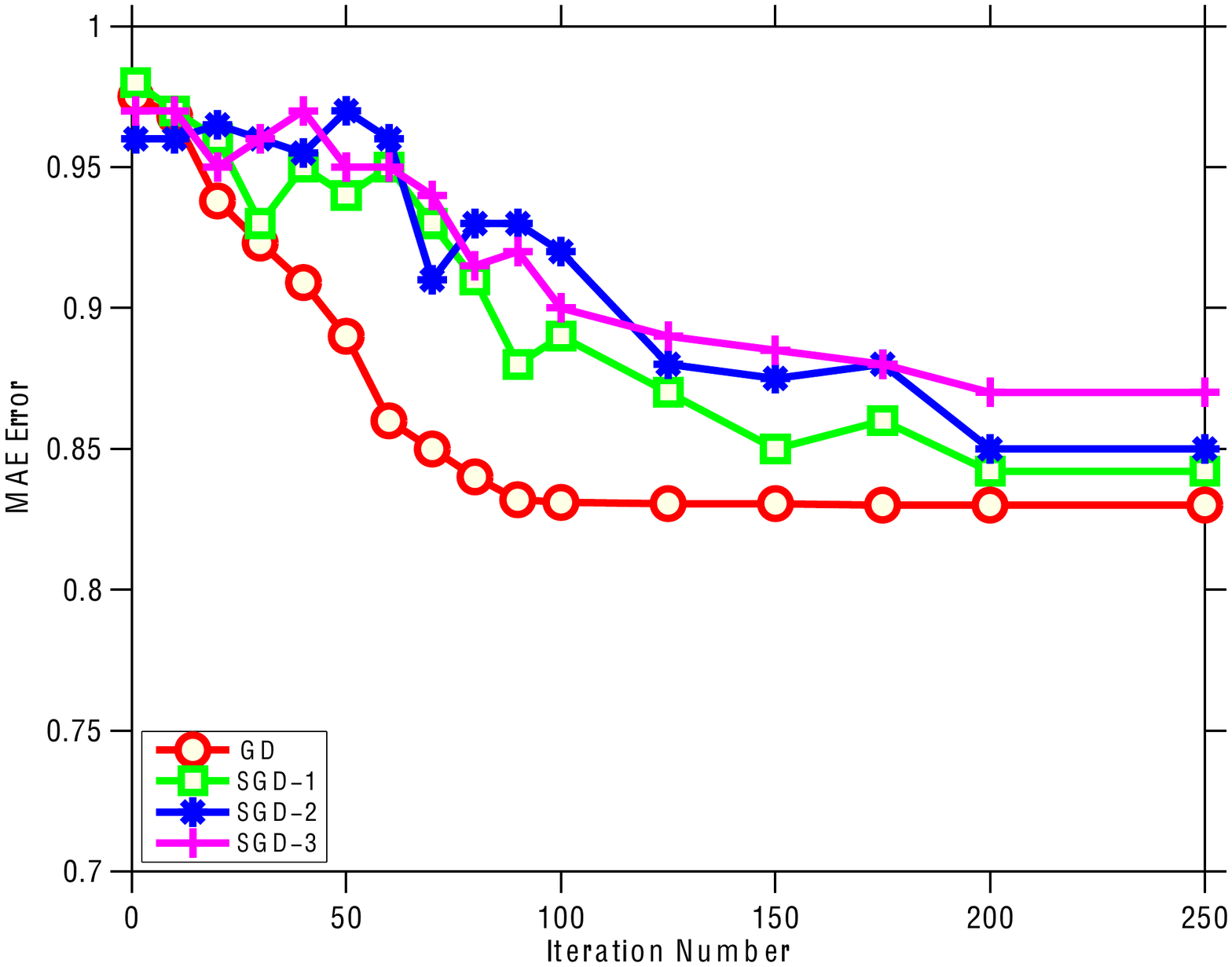}}\hspace*{-0.1in}
\subfigure[$70 \%$ of Training Data]{\includegraphics[scale=0.35]{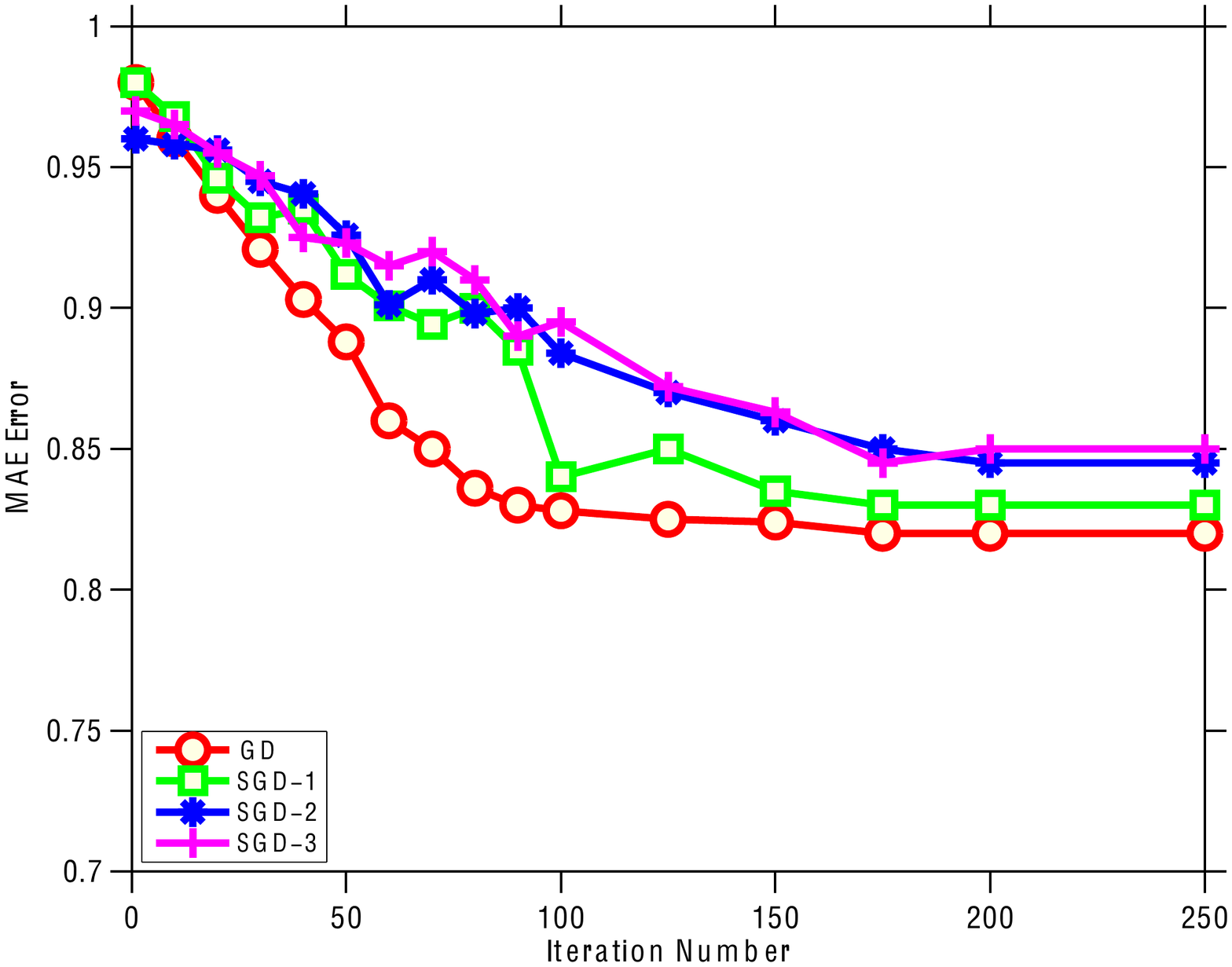}}
\centering
\subfigure[$80 \%$ of Training Data]{\includegraphics[scale=0.35]{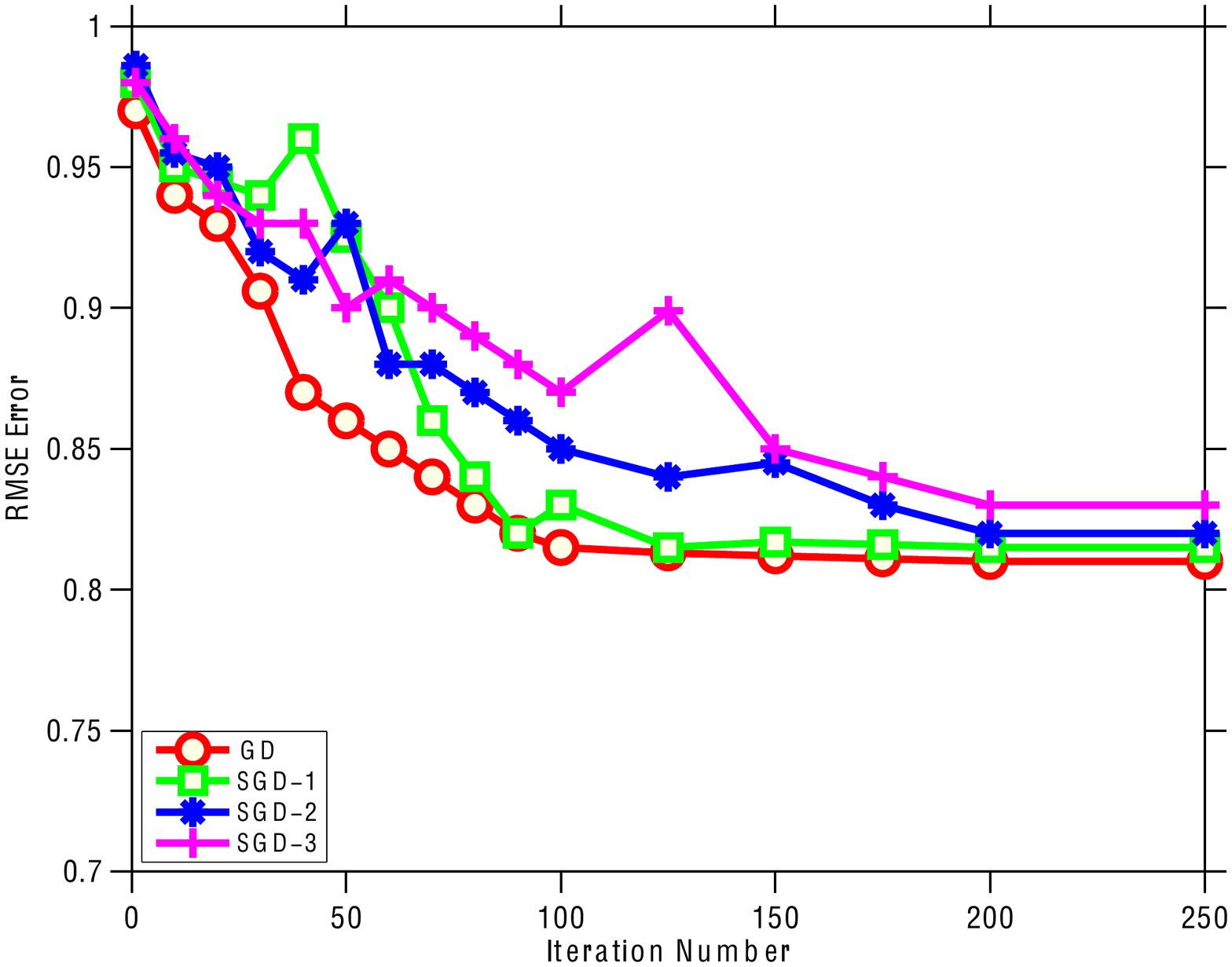}}\hspace*{-0.1in}
\subfigure[$90 \%$ of Training Data]{\includegraphics[scale=0.35]{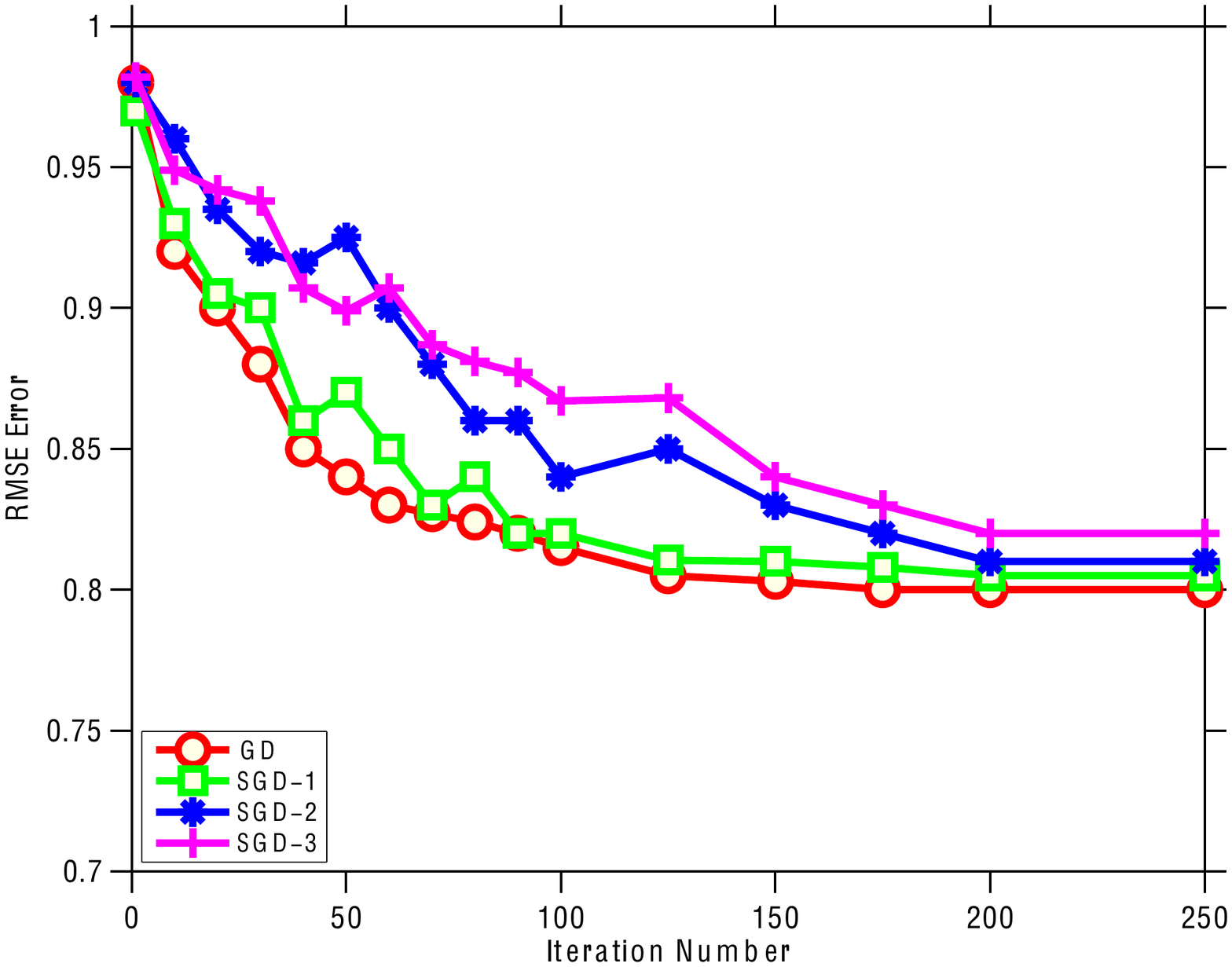}}
\caption{Comparison of accuracy of prediction in terms of MAE with GD and SGD with three varied batch sizes.}
\label{fig:rme}
\end{figure}
\subsection{On the Impact of Batch Size in Stochastic Optimization}
As mentioned earlier in the paper, directly solving the optimization problem in~(\ref{eqn:hing-euc}) using full gradient descent  method requires to go through all the triplets in the constraint set $\Omega_{\Sb}$ which could be computationally expensive due to the huge number of triplets  in  $\Omega_{\Sb}$. To overcome this efficiency problem, one can turn to stochastic gradient scent  method which tries to generate unbiased estimates of the gradient at each iteration in a much cheeper way by sampling a subset of triplets from $\Omega_{\Sb}$.  

To accomplish this goal, we perform gradient descent and stochastic gradient descent to solve the optimization problem in~(\ref{eqn:hing-euc}) to find the matrices  $\Ub$ and $\Vb$ following the updating equations derived in~(\ref{eqn:gradU}) and~(\ref{eqn:gradV}).  At each iteration $t$, the currently learned matrices $\Ub_t$ and $\Vb_t$  are  used to predict the ratings in the test set. In particular, at each iteration, we evaluate the RMSE and MAE on the test set, and terminate training once the RMSE and MAE starts increasing, or the maximum number of iterations  is reached.   We run the algorithm with latent vectors of dimension $k = 10$.

We compare the computational efficiency between proposed algorithm with  GD and  mini-batch  SGD with different batch sizes. We note that the GD updating rule can be considered as min-batch SGD where the batch size $B$ is deterministically set to be $B = |\Omega_{\Sb}|$ and simple SGD can be considered as mini-batch SGD with $B=1$. \ct{We remark that in contrast to GD method which uses all the  triplets in $\Omega_{\Sb}$ for gradient computation at each iteration, for SGD method due to uniform sampling over all tuples in $\Omega_{\Sb}$, some of the tuples may be  used more than once and some of the tuples might never been used for \textit{gradient} computation.  }

Figures~\ref{fig:rmse} and~\ref{fig:rme} show the convergence rate of four different updating rules in terms of the number of iterations $t$ for two different measures $\mbox{RMSE}$ and $\mbox{RME}$, respectively. The first algorithm denoted by GD runs the simple full gradient descent iteratively to optimize the objective.  The other three algorithms named SGD1, SGD2, and SGD3 in the figures use the batch sizes of $B = 0.1* |\Omega_{\Sb}|$, $B = 0.2* |\Omega_{S}|$, and $B = 0.3* |\Omega_{\Sb}|$, respectively. \ct{In our experiments, due to very slow convergence of the basic SGD method with $B =1$ in comparison to other fours methods, we simply exclude  its result from the discussion. }

\ct{In terms of accuracy of predictions, from both Figures~\ref{fig:rmse} and~\ref{fig:rme}, we can conclude that the GD has the best convergence and SGD3 has the worst convergence in all settings.} This is because, although all of the four algorithms use an unbiased estimate of the true gradient to update the solution at each iteration, but the variance of each stochastic gradient is proportional to the size of the batch size $B$. \ct{Therefore, for larger values of $B$, the variance of stochastic gradients is smaller and the algorithm convergences faster, but, for smaller values of $B$ the algorithm suffers from high variance in stochastic gradients and convergences slowly. We emphasize that this comparison holds for iteration complexity which is different from the computational complexity (running time) of individual iterations. More specifically, each iteration of GD requires $|\Omega_{\Sb}|$ gradient computations, while for SGD  we only need to perform  $B \ll |\Omega_{\Sb}|$ gradient computations. In summary, SGD has lightweight iteration but requires more iterations to converge. In contrast, GD takes expensive steps in much less number of iterations. From Figures~\ref{fig:rmse} and~\ref{fig:rme}, it is noticeable that although a large number of iterations is usually needed to obtain a solution of desirable accuracy using SGD, the lightweight computation per iteration makes SGD attractive for the optimization problem in~(\ref{eqn:hing-euc}) for large number of users.}  We also not that for the GD method, the error is a monotonically decreasing function it terms of number of iterations  $t$, but for the SGD based  methods this does not hold. This is because although SGD algorithm is guaranteed to converge to an optimal solution (at least in expectation), but there is no guarantee that the stochastic gradients provide a descent direction for the objective at each iteration due to the noise in computing  gradients. As a result, for few iterations we can see that the objective increases but finally it convergences as expected.

\section{Conclusions and Future Works}\label{sec:conc-future}
In this paper, we have made a progress  towards making distrust information beneficial in social recommendation problem. In particular, we have proposed a  framework based on the matrix factorization which is able to incorporate both trust and distrust relationships between users in factorization algorithm.   We experimentally investigated the potential of distrust as a side information to overcome the  data sparsity and cold-start problems in traditional recommender systems. In summary, our results showed that more accurate recommendations can be obtained by incorporating  distrust relations, indicating that  distrust information can indeed be beneficial for the recommendation process.

This work leaves few directions, both  theoretically and empirically, as future work.  From an empirical point of view, it would be interesting to extend our model for weighted social trust and distrust relations. One challenge in this direction is that, as far as we know, there is no publicly available data set that includes weighted (gradual) trust and distrust information.  Also, the   experimental results we have conducted on the consistency of social relations with rating information  hint at a number of potential enhancements  in future work. In particular, it would be interesting to further examine the correlation between implicit and explicit distrust information. An important challenge in this direction is to  develop better metrics to measure the implicit trust between users as the simple metrics such as Pearson  correlation coefficient seem to be insufficient.  Furthermore, since we only consider the distrust between users, it would be easy to generalize our model in the same way to incorporate dissimilarity between items and investigate how it works in practice.  Also, our preliminary results indicated that hinge loss almost performs better than the exponential loss, but from the optimization viewpoint, the exponential loss is more attractive due to its smoothness. So, an interesting direction would be to use a smoothed version of the hinge loss to gain from both optimization efficiency and algorithmic accuracy.


\bibliographystyle{plain}
\bibliography{social_recommendation}







\end{document}